\documentclass[a4paper,12pt]{article}
\usepackage[paper=letterpaper,margin=1in]{geometry}

\usepackage{amsmath,amssymb,amsfonts,epsfig,cite,setspace,bigstrut}
\pdfoutput=1

\begin{document}


\vskip 0.25in

\newcommand{\todo}[1]{{\bf ?????!!!! #1 ?????!!!!}\marginpar{$\Longleftarrow$}}
\newcommand{\fref}[1]{Figure~\ref{#1}}
\newcommand{\tref}[1]{Table~\ref{#1}}
\newcommand{\sref}[1]{\S~\ref{#1}}
\newcommand{\nn}{\nonumber}
\newcommand{\tr}{\mathop{\rm Tr}}
\newcommand{\comment}[1]{}

\newcommand{\cM}{{\cal M}}
\newcommand{\cW}{{\cal W}}
\newcommand{\cN}{{\cal N}}
\newcommand{\cH}{{\cal H}}
\newcommand{\cK}{{\cal K}}
\newcommand{\cZ}{{\cal Z}}
\newcommand{\cO}{{\cal O}}
\newcommand{\cB}{{\cal B}}
\newcommand{\cC}{{\cal C}}
\newcommand{\cD}{{\cal D}}
\newcommand{\cE}{{\cal E}}
\newcommand{\cF}{{\cal F}}
\newcommand{\cX}{{\cal X}}
\newcommand{\IA}{\mathbb{A}}
\newcommand{\IP}{\mathbb{P}}
\newcommand{\IQ}{\mathbb{Q}}
\newcommand{\IH}{\mathbb{H}}
\newcommand{\IR}{\mathbb{R}}
\newcommand{\IC}{\mathbb{C}}
\newcommand{\IF}{\mathbb{F}}
\newcommand{\IV}{\mathbb{V}}
\newcommand{\II}{\mathbb{I}}
\newcommand{\IZ}{\mathbb{Z}}
\newcommand{\re}{{\rm Re}}
\newcommand{\im}{{\rm Im}}
\newcommand{\li}{{\rm Li}}

\newcommand{\CA}{\mathbb A}
\newcommand{\CP}{\mathbb P}
\newcommand{\tmat}[1]{{\tiny \left(\begin{matrix} #1 \end{matrix}\right)}}
\newcommand{\mat}[1]{\left(\begin{matrix} #1 \end{matrix}\right)}
\newcommand{\diff}[2]{\frac{\partial #1}{\partial #2}}
\newcommand{\gen}[1]{\langle #1 \rangle}

\newcommand{\drawsquare}[2]{\hbox{%
\rule{#2pt}{#1pt}\hskip-#2pt
\rule{#1pt}{#2pt}\hskip-#1pt
\rule[#1pt]{#1pt}{#2pt}}\rule[#1pt]{#2pt}{#2pt}\hskip-#2pt
\rule{#2pt}{#1pt}}
\newcommand{\fund}{\raisebox{-.5pt}{\drawsquare{6.5}{0.4}}}
\newcommand{\antifund}{\overline{\fund}}

\newtheorem{theorem}{\bf THEOREM}
\def\thetheorem{\thesection.\arabic{theorem}}
\newtheorem{proposition}{\bf PROPOSITION}
\def\thetheorem{\thesection.\arabic{proposition}}
\newtheorem{observation}{\bf OBSERVATION}
\def\thetheorem{\thesection.\arabic{observation}}

\def\theequation{\thesection.\arabic{equation}}
\newcommand{\setall}{\setcounter{equation}{0}
        \setcounter{theorem}{0}}
\newcommand{\setequation}{\setcounter{equation}{0}}
\renewcommand{\thefootnote}{\fnsymbol{footnote}}

~\\
\vskip 1cm

\centerline{{\Large \bf $\cN=2$ Gauge Theories: Congruence Subgroups,}}
~\\
\centerline{{\Large \bf Coset Graphs and Modular Surfaces}}
\medskip

\vspace{.4cm}

\centerline{
{\large Yang-Hui He}$^1$ \&
{\large John McKay}$^2$
}
\vspace*{3.0ex}

\begin{center}
{\it
{\small
{${}^{1}$ Department of Mathematics, City University, London,\\
Northampton Square, London EC1V 0HB, UK;\\
School of Physics, NanKai University, Tianjin, 300071, P.R.~China;\\
Merton College, University of Oxford, OX14JD, UK\\
\qquad hey@maths.ox.ac.uk\\
}
\vspace*{1.5ex}
{${}^{2}$ 
Department of Mathematics and Statistics,\\
Concordia University, 1455 de Maisonneuve Blvd.~West,\\
Montreal, Quebec, H3G 1M8, Canada\\
\qquad mckay@encs.concordia.ca
}
}}
\end{center}

\vspace*{4.0ex}
\centerline{\textbf{Abstract}} \bigskip
We establish a correspondence between generalized quiver gauge theories in four dimensions and congruence subgroups of the modular group, hinging upon the trivalent graphs which arise in both.
The gauge theories and the graphs are enumerated and their numbers are compared.
The correspondence is particularly striking for genus zero torsion-free congruence subgroups as exemplified by those which arise in Moonshine.
We analyze in detail the case of index 24, where modular elliptic K3 surfaces emerge: here, the elliptic $j$-invariants can be recast as dessins d'enfant which dictate the Seiberg-Witten curves.
\newpage

\tableofcontents

\section{Introduction and Summary}
\setcounter{footnote}{0}
The modular group holds a central place in modern mathematics, vital to such diverse subjects as function theory, algebraic geometry, finite group theory and number theory.
The emergence of the modularity of $L$-functions of elliptic curves in the celebrated Taniyama-Shimura-Wiles theorem, and the encoding of the dimension of the representations of the Monster sporadic group into the $j$-invariant by Moonshine are but two manifestations of its unifying power.
In parallel, the modular group has also become an indispensible component of modern theoretical physics, especially in gauge and string theories.
The imposition of modular invariance on partition functions in conformal field theory and the organization of the spectra and couplings of supersymmetric theories by strong-weak S-duality clearly exemplify its physical significance.

This ubiquity of the modular group constitutes the corner-stone of our subconscious as we approach our problem from the point of view of graphs, $\cN=2$ gauge theory and Calabi-Yau geometry.
Our starting point will be a perhaps less well-known fact about the modular group $\Gamma := PSL(2;\IZ)$, that its Cayley graph is an infinite free trivalent tree, but with each node replaced by an oriented triangle.
This fact by itself may not be immediately inspiring, however, when we consider subgroups of finite index within $\Gamma$ more non-trivial graphs emerge.

Indeed, if we quotient the graph by action of the subgroup, we arrive at the so-called {\bf Schreier-Cayley coset} graph, which remains to be trivalent, but is now finite. 
In fact, the coset graph is finite if and only if the subgroup stabilizing a node is of finite index; indeed,
all connected finite trivalent graphs can be realized in this way.
Thus, we can translate the investigation of these graphs - also known as cubic graphs - to that of finite index subgroups of the modular group, a well studied subject in number theory.

One of the most important class of subgroups is the the so-called genus zero subgroups, which are those which produce, up to cusps, the Riemann sphere when quotienting the upper half plane.
In Moonshine, the encoding of the dimensions of the irreducible representations of the Monster Group by the elliptic $j$-invariant is only the tip of the iceberg since the dimensions are the characters for the identity representation.
In general, the McKay-Thompson series for the characters of an arbitrary element of the Monster group yield principal moduli precisely for the genus zero subgroups.
Naturally, a classification of such groups (in addition to torsion-freeness so as to give finite index) was in demand, a question nicely settled in \cite{classSebbar}, which showed that they are very rare.

The Schreier-Cayley coset graphs for the genus zero, torsion-free congruence subgroups from the classification were subsequently studied in \cite{mckaysebbar,sebbar}, a distinguished set of trivalent finite graphs thus emerged, each corresponding to a modular curve which is a Riemann sphere $\IP^1$, which we denote as $\IP^1_C$.
Further structure emerges when one considers elliptic fibrations over these special $\IP^1_C$'s, giving us complex algebraic surfaces; these are the so-called semi-stable {\bf modular elliptic surfaces} the nature of whose fibration is controlled by the cusps.
In particular, there are nine members in the list which have the distinction of being index 24 subgroups of the modular group and the associated modular elliptic surfaces are K3.
This Calabi-Yau avatar of these groups is certainly inspiring, both from the geometric and the arithmetic points of view.
Our first task is then to describe these above matters in detail in \sref{s:math}.

The point d'appui of our discussion hinges on a set of finite trivalent graphs.
This instantly reminds us of the vital appearance of such graphs in a seemingly unrelated subject, that of gauge theories in four space-time dimensions.
Over the last two years, a set of $\cN=2$ gauge theories have been distinguished by Gaiotto \cite{Gaiotto:2009we}. They have interesting duality behaviour amongst themselves under S-duality and possess a systematic construction for their {\bf Seiberg-Witten curves}.
The string-theoretic realization is very geometrical: take a Riemann surface $\Sigma_G$ of genus $g$ with $e$ punctures, dubbed the {\bf Gaiotto curve}, and wrap $N$ coincident parallel M5-branes over it.
The world-volume theory on the M5-branes is one with a product $SU(N)$ groups, whose Seiberg-Witten curve is exactly an $N$-fold cover $\Sigma_G$.

Let us, for convenience, henceforth take $N=2$ and what we have is a theory with $SU(2)^{3g-3+e}$ gauge group, with $2g-2+e$ half-hypermultiplets transforming under the tri-fundamental representation of $SU(2)^3$, together with $e$ massive flavours globally charged under the $SU(2)$ groups.
Such a theory can be succinctly encoded into a {\bf skeleton diagram} \cite{Hanany:2010qu} which generalizes a quiver diagram.
This is a trivalent graph whose edges correspond to $SU(2)$ groups and nodes are the tri-fundamental matter; we also allow external legs of valency 1 corresponding to the flavours.
Hence. we have a graph with $g$ closed circuits, $3g-3+e$ edges, $2g-2+e$ nodes, and $e$ external lines.
The diagram actually constitutes the spine of the am{\oe}ba projection of the Gaiotto curve and hence captures the genus and number of punctures on $\Sigma_G$ and, in turn, the Seiberg-Witten curve which is a two-fold covering of $\Sigma_G$.
Finally, the moduli space of vacua of the theory is specified by the topology $(g,e)$ of the graph.
Detailing these $\cN=2$ gauge theories will be the subject of \sref{s:phys}.

The parallel vein between the above two strands, one from mathematics and the other from physics, centering upon trivalent graphs and Riemann surfaces is certainly suggestive and compels us to proceed further and strengthen the connection.
We focus on the nine elliptic K3 surfaces corresponding to index 24 subgroups, the graphs for which are drawn in \fref{f:graph24}.
These are readily re-interpreted as skeleton diagrams for $\cN=2$ theories with $SU(2)^{12}$ gauge group, 8 half-hypermultiplets and no flavours.
Remarkably, the vacuum moduli space (Kibble branch) of these gauge theories have been computed \cite{Hanany:2010qu} to be exactly local K3 surfaces.

The precise equations for the K3 surfaces as semi-stable extremal fibrations over $\IP^1_C$ were worked out in \cite{LY,TopYui} and the $j$-invariants for all genus zero subgroups, in \cite{mckaysebbar}.
We recompute them here in a slightly different form, emphasizing the inter-relations amongst the groups and correcting some misprints sprinkled in the literature, and summarize them in \tref{t:weier}.

The $j$-invariants of these surfaces \cite{MP,BM}, as elliptic curves depending on a base parameter which is the projective coordinate on the base $\IP^1_C$, turn out to be rational maps ramified over a target $\IP^1$ only at 0, 1 and $\infty$.
Such a map is called a {\bf Belyi map} and has a presentation as a finite connected bipartite graph known as a {\bf dessin d'enfant}, which in its original motivation by Grothendieck controls the number field and associated Galois group action over which the pre-image is defined as an algebraic curve.
Here, all our K3 surfaces and subsequent $j$-maps, can be defined over $\IQ$.
The dessins are clean - in that the ramification over 1 is 2 - and are drawn in \fref{f:dessin}.
They turn out to be precisely the Schreier-Cayley coset graphs, decorated with black/white nodes.

What is more pertinent for our purposes is that these dessins encode the Seiberg-Witten curves of $\cN=2$ gauge theories \cite{hep-th/0611082}.
In particular, a clean Belyi map from $\IP^1$ to $\IP^1$ can always be considered as the non-trivial part of a hyper-elliptic curve which is the Seiberg-Witten curve and the nature of the roots, i.e., the factorization, is dictated by the Belyi map and, graphically, by the dessin.
We apply this prescription in \sref{s:SW-dessin}, using the fact that the Seiberg-Witten curves can be retrieved as a ramified cover of the Gaiotto curves, and find that the $j$-invariant for our Belyi maps for the nine modular elliptic K3 surfaces can be chosen to give a Seiberg-Witten curve for $\cN=2$ gauge theories, at specific points in moduli space of couplings and parameters. 

In this way, we close the two directions undertaken at the beginning of our exploration, initiated by the trivalent graphs.
For amusement, in \sref{s:SW-revisit}, we also include the analysis of the elliptic modular K3 surfaces, treating them as Seiberg-Witten curves of an $SU(2)$ theory, identifying the base coordinate of $\IP^1_C$ with the $u$-plane and tuning various mass parameters.

Fortified by this correspondence, it is tempting to conceive of a general recipe (for reference, we include an enumeration of the total possible such theories in \sref{s:count} to give an idea of the plethora of possibilities). We start by taking any $\cN=2$ Gaiotto theory, which is specified by a Riemann surface $\Sigma_G$ with possible punctures, the spine of whose am{\oe}ba projection is a finite trivalent graph, possibly with external lines.
Now, any trivalent graph is a Schreier-Cayley coset graph of a subgroup of the modular group\footnote{
The external edges are univalent and give rise to complications.
We can either restrict to pure $\cN=2$ theories without flavours, or include them in the coset graph, which are called ``spikes'' in \cite{mckaysebbar}.
All the examples in this paper are without spikes.
} so we can associate\footnote{
By Moonshine, we can therefore also associate appropriate representations of the Monster group to the given gauge theory; this is clearly another direction to pursue.
} a Riemann surface $\Sigma_C$ formed by quotienting the upper half plane by this modular subgroup, adjoining cusps as necessary.
It then remains to investigate the relation between $\Sigma_G$ and $\Sigma_C$.

In this paper, we have uncovered many intricate relations for the nine index 24 congruence subgroups: here $\Sigma_C$ are spheres over which an elliptic fibration gives extremal modular K3 surfaces; the $j$-invariants for these are clean Belyi maps and give back possible Seiberg-Witten curves which are double covers of $\Sigma_G$ for the original theory.
Clearly, many of the special properties of K3 surfaces are inherent in our correspondence.
However, given the multitude of congruence subgroups, trivalent coset graphs, and $\cN=2$ gauge theories, how far this story may extend is unquestionably an interesting venue for future investigations.

\section{Dramatis Person\ae}
\setcounter{footnote}{0}
In this section, we introduce the two sides of our correspondence, and meticulously set the nomenclature as well as attempt to include all elementary definitions in a manner as self-contained as possible so as to facilitate physicists and mathematicians alike.
First, we describe standard subgroups of the modular group $\Gamma$ and distinguish the torsion-free, genus zero congruence subgroups.
Along the way, we discuss the Cayley graphs for the modular groups and the Schreier-Cayley coset graphs for the subgroups.
We show how these particular groups give rise to modular elliptic surfaces and why those of index 24 are special.

Second, we turn to the physics and present a wide class of $\cN=2$ four-dimensional gauge theories which have of late attracted much attention.
This is due to the fact that they have interesting classical moduli spaces of vacua, that their Seiberg-Witten curves are succinctly captured by Riemann surfaces with punctures, and that the theories can be graphically encoded into so-called skeleton diagrams which are trivalent.
Thus prepared, we will, in the next section, combine these two seemingly unrelated strands of thought and weave a intriguing story.

\subsection{The Modular Group and Cayley Graphs}\label{s:math}
We begin with some standard preliminary definitions to set notation.
It is well known that the {\it modular group} $\Gamma = PSL(2; \IZ)$ of linear fractional transformations $z \to \frac{a z + b}{c z + d}$, with $a,b,c,d \in \IZ$ and $\det \left( \begin{matrix} a & b \\ c & d \end{matrix} \right) =1$ is finitely generated by $S : z \to -1/z$ and $T : z \to z+1$, such that
\begin{equation}
\Gamma := PSL(2;\IZ) \simeq \langle S, T  \, \left| \, S^2 = (ST)^3 = I \rangle \right. 
\ .
\end{equation}
Calling $x$ the element of order 2 and $y$ the element of order 3, we see that $\Gamma$ is the free product of the cyclic groups $C_2 = \gen{x | x^2 = I}$ and $C_3 = \gen{y | y^3 = I}$.
That is, $\Gamma \simeq C_2 \star C_3$.
Thus written, one can easily construct the {\bf Cayley Graph} for $\Gamma$.
We recall that this is a directed graph whose every node corresponds to an element of a finitely generated group and such that there is an arrow from node $g$ to $g'$ if there exists an element $h$ in the group such that $g' = hg$.
Given the $C_2 \star C_3$ structure of $\Gamma$, we see that $x$ will serve as a bi-directional arrow whilst $y$ will give rise to an oriented triangle, namely, a directed triangular closed circuit\footnote{
We will adhere to the graph theoretic notation that a ``loop'' refers to a self-adjoining arrow to a single node and a ``closed circuit'' that being formed by traversing edges, returning to a node. 
}.
Therefore, the Cayley graph of the modular group is an infinite free trivalent tree with each node replaced by an oriented 3-cycle.
Note that we may, without loss of generality,
take all nodes to be positively oriented. 
This is not necesarily so for finite graphs.
The reader is referred to, for example, Figure 1(c) of \cite{CayleyGamma}.

\subsubsection{Coset Graphs}
The Cayley graph of $\Gamma$ can be folded to generate finite trivalent (or ``cubic'') graphs which will constitute the subject of this note.
For a subgroup $G \in PSL(2;\IZ)$ of index $\mu$, we can decompose the modular group into the (right) cosets $G g_i$ of $G$ as 
\begin{equation}
PSL(2;\IZ) \simeq \bigcup\limits_{i=1}^\mu G g_i \ , 
\end{equation}
so that our generators $x$ and $y$ act by permuting the nodes, which now correspond to cosets.
The result is a coset graph with $\mu$ nodes and a folded version of the Cayley graph of the full modular group.
We will call it a Schreier-Cayley coset graph and
it remains, in particular, to be trivalent, with bi-directional edges for $x$ and oriented 3-cycles for $y$.
In fact, the converse is true: {\it any} finite cubic graph is a realization of a Schreier-Cayley coset graph of a subgroup of the modular group.

\begin{figure}[!h!t!b]
\centerline{
\includegraphics[trim=0mm 0mm 0mm 0mm, clip, width=5.0in]{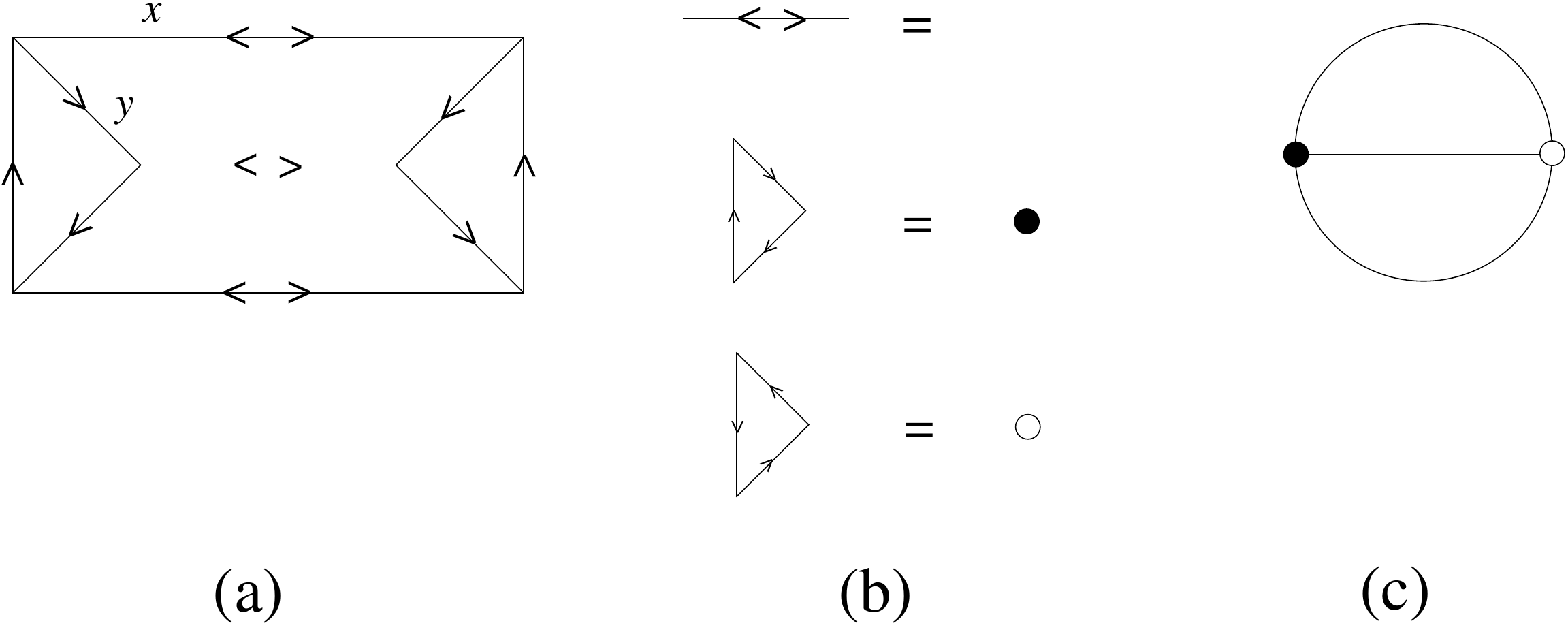}
}
\caption{{\sf {\small
The Schreier-Cayley coset graph, drawn in (a), of an index 6 subgroup of the modular group, shrunken, using the rules of (b), and represented in (c).
}}
\label{f:coset-eg}}
\end{figure}

As an example, consider the diagram in \fref{f:coset-eg}.
Here $\mu = 6$ and the subgroup is thus one of index 6 (it turns out to be $\Gamma(2)$, as we will discuss shortly). In part (a) of the figure, we have drawn the coset graph as a di-graph, marking the arrows explicitly: the bi-directional edge corresponds to $x$ while the directed tri-cycle is generated by $y$. In part (b), following the notation of \cite{sebbar}, we will henceforth, for the sake of brevity, denote each such bi-directional edge by a simple edge and shrink each tri-cycle to a node. 
Now, for such coset graphs, we clearly have choice of orientation for the tri-cycle: positive or negative.
We will, as convention, take the former to be shrunk to a black node, and the latter, to a white node\footnote{
Though the resulting graph may be reminiscent of dimer models which have also appeared in the gauge theory context (cf.~\cite{Feng:2005gw}), we must emphasize that our graphs have no necessity to be bipartite and nodes of the same colour can well be adjacent.
We will, however, turn to these bipartite graphs later.
}.
Thus, the coset graph in (a) is succinctly represented as in part (c).
If we ignore the colour, then this diagram has in a different context, as we will soon see, been called ``the sunset'' or ``Yin-Yang'' diagram.
Indeed, the shrinking process preserve the trivalent nature of our graphs of concern.

Furthermore, 
it turns out that for the particular subgroups and coset graphs we are considering in this paper, the triangles in the graphs can all be chosen to have the same orientation, say positive.
Therefore, henceforth, we will no long mark the colour of the nodes of our coset graphs and we do have standard trivalent graphs to study.

\subsubsection{Torsion-Free and Genus Zero Congruence Subgroups}
Let us first introduce some standard notation. Recalling that the modular group is itself $SL(2;\IZ) / \{\pm I \}$, we have the following subgroups:
\begin{itemize}
\item Principal Congruence subgroup
\[
\Gamma(m) := \{A \in SL(2;\IZ)  \, \left| \, A \equiv \pm I \bmod m \right.\} 
/ \{\pm I \} \ ;
\]
\item Congruence subgroup of level $m$: subgroup of $\Gamma$ containing $\Gamma(m)$ but not any $\Gamma(n)$ for $n < m$;
\item Unipotent matrices
\[
\Gamma_1(m) := \left\{A \in SL(2;\IZ)  \, \left| \, A \equiv \pm \left( \begin{matrix}
1 & b \\ 0 & 1
\end{matrix} \right) \bmod m \right. \right\} / \{\pm I \} \ ;
\]
\item Upper triangular matrices
\[
\Gamma_0(m) := \left\{ \left. \left( \begin{matrix}
a & b \\ c & d
\end{matrix} \right) \in \Gamma  \, \right| \, c \equiv 0 \bmod m \right\} / \{\pm I \}
\ .
\]
\end{itemize}
Indeed, $\Gamma_0(m)$ and $\Gamma_1(m)$ are both examples of congruence subgroups at level $m$. Furthermore, $\Gamma(1) = \Gamma$ is understood to be the full modular group and we shall often use this notation; likewise, $\Gamma$ with subscripts and/or arguments will be used to denote some congruence subgroup.

Elements $A$ of $SL(2;\IC)$ fall into four cases:
\begin{itemize}
\item Parabolic: $\tr(A) = \pm 2$ for $A \ne \pm \II$; 
\item Elliptic: $\tr(A) \in \IR$ and $|\tr(A)| < 2$;
\item Hyperbolic: $\tr(A) \in \IR$ and $|\tr(A)| > 2$;
\item Loxodromic: $\tr(A) \not\in \IR$ \ ,
\end{itemize}
and the first three also apply to the modular group $PSL(2;\IZ) \subset SL(2;\IC)$.

In \cite{mckaysebbar,sebbar}, a particular family of subgroups of the modular group has been identified. These are the so-called {\bf torsion-free} and {\bf genus zero} congruence subgroups.
By torsion-free we mean that the subgroup contains no element, other than the identity, which is of finite order.
To explain genus zero, let us again recall some rudimentaries (cf.~e.g.~\cite{serre,Milne}).

We know that the modular group acts on the upper half-plane $\cH := \{ \tau \in \IC \; | \; \im(\tau) > 0 \}$ by linear fractional transformations.
It is also a well-known fact that ${\cal H}$ gives a compactification $\cH^*$  when adjoining so-called ${\it cusps}$, which are points on $\IR \cup \infty$ fixed under some parabolic element\footnote{It can be shown that such cusp points for $\Gamma(1)$, except $\infty$, are rational.}.
Subsequently, the quotient $\cH^* / \Gamma(1)$ is a compact Riemann surface of genus 0, that is, a sphere.
The natural question thus arises as to how this generalizes to the congruence subgroups.
It turns out that with the addition of appropriate cusp points, the extended upper half plane $\cH^*$ factored by various congruence subgroups will also be compact Riemann surfaces, possibly of higher genus.
Such a Riemann surface is called a {\bf modular curve}.
Specifically, using the Riemann-Hurwitz theorem, the genus of the modular curve can be shown to be:
\begin{equation}
g_C = 1 + \frac{\mu}{12} - \frac{\nu_2}{4} -  
    \frac{\nu_3}{3} - \frac{\nu_\infty}{2} \ ,
\end{equation}
where $\mu$ is the index of the subgroup, $\nu_2, \nu_3$ are the numbers of inequivalent elliptic points of orders 2 and 3 respectively and $\nu_\infty$, the number of inequivalent cusp points.

It turns out that genus 0 subgroups are very rare and are of a special importance in \textit{Monstrous Moonshine}.  Indeed, the groups give rise to modular curves whose function field has a single (transcendental) generator, such as the $j$-invariant for the $\Gamma(1)$ itself. This unique (up to modular transformation) function can be normalized to a \textit{principal modulus} \footnote{Introduced in \cite{cox}, and sometimes called ``hauptmodul'' in the literature.
}
whose $q$-expansion can be related to the characters of the so-called head representations of the Monster sporadic group.

We are interested in genus $g_C=0$ and torsion-free subgroups, thus we have that $g_C=\nu_2=\nu_3=0$, whence we have a further constraint that $\mu = 6 \nu_\infty - 12$, so that the index of all our subgroups of concern is a multiple of 6.
In fact, all but two (of index 6) are multiples of 12.
In \cite{classSebbar}, a complete classification of such subgroups is performed and they fall into only 33 conjugacy classes, all of index 6, 12, 24, 36, 48 or 60.
They are:
\begin{itemize}
\item $\Gamma(n)$ with $n=2,3,4,5$;
\item $\Gamma_0(n)$ with $n=4,6,8,9,12,16,18$;
\item $\Gamma_1(n)$ with $n=5,7,8,9,10,12$;
\item the intersections $\Gamma_0(a) \cap \Gamma(b)$ for $\{a,b\} =
\{4,2\}, \{3,2\}, \{8,2\}, \{2,3\}, \{25,5\}$;
$\Gamma_1(8)\cap \Gamma(2)$ and $\Gamma_0(16)\cap \Gamma_1(8)$
\item the congruence subgroups
\[
\Gamma(m; \frac{m}{d}, \epsilon, \chi) :=
\left\{ \left.
\pm \left(\begin{matrix}
1 + \frac{m}{\epsilon \chi} \alpha & d \; \beta\\
\frac{m}{\chi} \gamma & 1 + \frac{m}{\epsilon \chi} \delta
\end{matrix}\right) \, \right| \,
\gamma \equiv \alpha \bmod \chi 
\right\} \ ,
\]
with $(m,d,\epsilon, \chi) = (8,2,1,2), (12,2,1,2), (16,1,2,2),
(27,1,3,3), (8,4,1,2)$, \\
$(12,2,1,2), (16,2,2,2), (24,1,2,2), (32,1,4,2)$.
\end{itemize}
The coset graphs for these particular subgroups are studied in \cite{mckaysebbar}.

\subsubsection{Modular Elliptic Surfaces}\label{s:modSurface}
We now move on to a generalization of modular curves, viz., modular surfaces, first introduced in \cite{shioda} and to \cite{shioda-rev} we refer the readers for an excellent review.
It will turn out that promoting our modular curve to a surface gives us beautiful new structures, both geometrically and number-theoretically.

First, we extend the action of any subgroup $\Gamma$ of the modular group on $\cH$ to an action
\begin{equation}
\cH \times \IC \ni (\tau, z)
\longrightarrow \left( \gamma \tau, \frac{z + m \tau + n}{c \tau + d}
\right) \ ,
\end{equation}
for $\gamma = \mat{a & b \\ c & d} \in \Gamma$  and $(m,n) \in \IZ^2$.
Thus the quotient of $\cH \times \IC$ by the above automorphism defines a surface equipped with a morphism to the modular curve arising from the quotient of $\cH$ by $\tau \to \gamma \tau$.
The fiber over the image of this morphism to the modular curve is generically an elliptic curve corresponding to the lattice $\IZ \oplus \IZ \tau$.
What we have therefore is a complex surface which is an elliptic fibration\footnote{
This actually becomes a universal covering of the moduli space of the modular curve; we thank Alessio Corti and Ed Segal for pointing this out.
} over the modular curve, called the {\bf elliptic modular surface} associated to $\Gamma$.
The base, because our modular curves are genus zero, will be the Riemann sphere $\IP^1_C$. Hence, our modular surfaces will be elliptic fibrations over $\IP^1$, which is a well-studied subject.

The classification of elliptic fibration structures due to Kodaira subsequently dictates that the Euler characteristic of the modular surface is the index of the subgroup $\Gamma$, thereby giving the geometric genus $p_g = \frac{\mu}{12}-1$.
Consequently, Top and Yui \cite{TopYui} further show that the index 12 congruence subgroups are rational elliptic surfaces, the index 24 subgroups are elliptically fibred K3 surfaces and the remaining, surfaces \comment{of general type} with geometric genera 2,3,4;
they have developed explicit equations for these surfaces.
Moreover, the degeneration type of the elliptic fibres, according to Kodaira, is $I_n$ for all our surfaces. Such elliptic surfaces are called {\it semi-stable}.
These singular fibres all reside on the cusps of the base modular curve, and the integer $n$ for the singularity type $I_n$ is called the {\it cusp width}; the list of cusp widths is a characterizing feature for each surface.
This data will be of great use to us later.

We remark that the Euler number of all the modular surfaces of genus zero is one of 6, 12, 24, 36, 48 and 60, thus perhaps the cases of 12, 24 and 60 are of some distinction since they are the order of the $A_4$, $S_4$ and $A_5$ finite groups.
Their central extensions are the binary polyhedral groups which are subgroups of $SU(2)$ and are related to $E_{6,7,8}$ groups in the ADE classification by the McKay Correspondence.
Here, these are, respectively, rational elliptic surfaces, elliptic K3 surfaces and extremal surfaces of N\'eron-Severi group of rank equal to the Hodge number $h^{1,1} = 50$. 
The K3 surfaces will be our chief concern later.

Here concludes the introduction to the congruence subgroup, coset graph and modular surfaces side of our story.
In the following subsection, we will turn to the physics of gauge theories.

\subsection{$\cN = 2$ Gauge Theories and Dualities}\label{s:phys}
Now, let us introduce the other side of the correspondence.
In \cite{Gaiotto:2009we}, Gaiotto found a new and interesting class of $\cN=2$ supersymmetric gauge theories in four dimensions, obtainable from the wrapping of M5-branes over Riemann surfaces. 
Indeed, it is a famous fact that the S-duality group for $\cN=2$ $SU(N)$ gauge theories is precisely the modular group, acting by linear fractional transformations on the complexified gauge coupling $\tau = \frac{\theta}{\pi} + \frac{2 \pi i}{g_{YM}^2}$, incorporating the Yang-Mills coupling $g$ and the theta-angle $\theta$.

Following \cite{Hanany:2010qu}, let us focus on the case where the gauge group is only products of $SU(2)$ factors. In this case, we can unambiguously represent the relevant gauge theories as so-called {\bf skeleton diagrams}, consisting
of lines and trivalent nodes, where a line represents an $SU(2)$ gauge group and
a trivalent node represents a matter field in the tri-fundamental representation of
$SU(2)^3$. Hence, these diagrams can be seen as generalizations of the more familiar \textit{quiver diagrams} which have arisen both in the representation theoretical \cite{ADE} and gauge theoretic context (e.g.~\cite{Hanany:1998sd}).
Indeed, whereas fields charged under {\it two} $SU$-factors, being the fundamental under one and the anti-fundamental of another, readily afford descriptions in terms of arrows in a quiver, fields charged under more than two factors, as in our present case, require encodings beyond a quiver diagram.

Our skeleton diagrams are straight-forward: they give rise to an infinite class of $\cN=2$ gauge theories, having each line representing an $SU(2)$ gauge group with its length inversely proportional\footnote{Therefore, any line of infinite length would give zero coupling and the associated $SU(2)$ becomes a global symmetry. As mentioned in the introduction, such situations with ``spikes'' will actually not arise in our correpondence with the modular coset graphs and is thus not our present concern.
} to its gauge coupling $g_{YM}^2$ and each trivalent node representing a half-hypermultiplet $Q_{\alpha\beta\gamma}$ transforming under the tri-fundamental $(\fund,\fund,\fund)$ representation of $SU(2) \times SU(2) \times SU(2)$, with $\alpha, \beta, \gamma = 1,2$ indexing each of the $SU(2)$ factor.
Because $\cN=2$ supersymmetry is large enough to have its matter content determine the interactions completely, each skeleton diagram thus defines a unique $(3+1)$-dimensional Lagrangian for the gauge theory.
As an example, we can turn back to part (c) of \fref{f:coset-eg}.
This can now be interpreted as an $SU(2)^3$ theory with 2 half-hypermultiplets, charged as tri-fundamentals.

The string-theoretic realization of these theories is in terms of a stack of M5-branes (here two M5-branes in order to give the $SU(2)$) wrapping a Riemann surface $\Sigma_G$ of genus $g$ with $e$ punctures.
More precisely, 
consider M-theory in 11-dimensions with coordinates $x^{0, \ldots, 10}$ with $x^{7,8,9} = 0$ fixed and $x^{0,1,2,3}$ the coordinates of our four-dimensional world $\IR^4_{x^{0,1,2,3}}$. 
Of the remaining 4 directions $Q_{x^{4,5,6,10}} \simeq \IR^3 \times \IR$, define a complex structure $v = x^4 + i x^5$ and $t = \exp(-(x^6 + i x^{10})/R_{IIA})$ (so that the $x^{10}$ direction indeed becomes periodic when compactifying to type IIA string theory on a circle of radius $R_{IIA}$), and define a Riemann surface $\Sigma_{G} = \{F(v,t) = 0\} \subset Q$ over which the M5-brane can wrap.
In the type IIA perspective this corresponds to $n+1$ NS5 branes occupying $x^{0,1,\ldots,5}$ and placed in parallel at fixed values of $x^6$; moreover, between adjacent pairs of NS5-branes are stretched stacks of D4-branes.
The variable names are chosen judiciously: the skeleton diagram of the theory to which $\Sigma_G$ corresponds is one whose topology, graphically, consists of $g$ independent closed circuits (to avoid confusion with previous discussion, we will avoid calling this the ``genus'' of the finite graph) and $e$ external (semi-infinite) legs.

Indeed, we can easily check that given a skeleton diagram specified by the pair $(g,e)$, the number of internal (finite) lines, hence the number of $SU(2)$ gauge group factors, is $3g-3+e$ while the number of nodes, hence the number of matter fields is the Euler character $2g-2+e$.

\subsubsection{The Moduli Space: Kibble Branch}
In supersymmetric gauge theories, the vacuum expectation values (VEV) of the scalar component fields of the multiplets parametrize the vacuum configurations of the field theory; this is called the (classical) vacuum {\bf moduli space}.
One could think of this as the space of minima to an effective potential, which is usually governed by the vanishing of a set of algebraic equations.
In other words, the moduli space is an affine \textit{algebraic variety}, realized in some complex space $\IC^k$ whose coordinates are the VEVs.
As a variety, the moduli space can often have non-trivial primary decomposition into components customarily called {\it branches}.

We follow the nomenclature of \cite{Hanany:2010qu} and focus on the {\bf Kibble Branch} parametrized by the VEVs of the gauge singlets of our hypermultiplet fields\footnote{In standard $SU(N)$ $\cN=2$ supersymmetric gauge theories, one usually has the {\it Higgs} and {\it Coulomb} branches of the moduli space.
The former is parametrized by the massless gauge singlets of the hypermultiplets, occurring where the gauge group is completely broken and the vector multiplet becomes massive via the Higgs mechanism. The latter is parametrized by the complex scalars in the vector multiplet, occurring when the gauge group is broken to some Abelian subgroup and the hypermultiplets generically becoming massive.
Here, for $g > 0$, where there is more than a single $SU(N)$ factor, the gauge group may not be completely broken on the Higgs branch of each factor and thus to avoid confusion, the authors of \cite{Hanany:2010qu} dub this quasi-Higgs branch after Tom Kibble. Of course, for $g=0$, $\cK$ is just the Higgs branch.
}, denoted as $\cK$.
A beautiful result from \cite{Hanany:2010qu} is that the Kibble branch of the moduli space is an algebraic variety such that
\begin{equation}\label{dimK}
\dim_{\IH} \cK = e + 1 \ ,
\end{equation}
where $\dim_{\IH}$ means the \textit{quaternionic} dimension, i.e., four times the real dimension or twice the complex dimension.
It is interesting to see that the result is independent of $g$.
A quick argument can proceed as follows.
Each trivalent node consists of 4 quaternionic degrees of fredom and there are $\chi = 2g-2+e$ thereof; generically on $\cK$, the $SU(2)^{3g-3+e}$ gauge group breaks to $U(1)^g$, hence $3(3g-3+e)-g$ number of broken generators.
Thus there is effectively, $4\chi - (3(3g-3+e)-g) = e+1$ quaternionic degrees of freedom.
One particular corollary to \eqref{dimK} is that in the absence of external (semi-infinite) legs, which will constitute all the graphs of our interest, especially in relation to the previous section on coset graphs, the Kibble branch is always a complex algebraic surface.
Furthermore, since these surfaces have quaternionic structure, they are in fact hyper-K\"ahler spaces. Hence, they have to be K3 surfaces, at least locally (since they are affine varieties).

As an example, let us once more turn to part (c) of \fref{f:coset-eg}.
The Kibble branch was computed in \cite{Hanany:2010qu} to be the local K3 surface prescribed by the ADE singularity  $\IC^2 / \hat{D}_3$ where $\hat{D}_3$ is the binary dihedral group of order 4, with the Hilbert series
$H(T) = \frac{1-T^8}{(1-T^2)(1-T^4)^2}$ so that we have a hypersurface in $\IC^3$ with one variable of weight 2, two of weight 4, obeying a single equation in degree 8.
Furthermore, this is shown to be the the same as that for the ``dumb-bell'' diagram consisting of two circles joined by a line, since they are both of $g=2$ and $e=0$.

In general, as shown in \S7.1 in loc.~cit., the Kibble branch of the moduli space is dependent only on the topology of the skeleton diagram, and is thus specified by the pair $(g,e)$, with $g$ both the number of closed circuits in the diagram and the genus of the Riemann surface $\Sigma_G$ which the M5-branes need to wrap in order to produce the gauge theory and $e$ the number of external legs as well as the number of punctures in $\Sigma_G$.
In particular, for $e=0$, the Kibble branch is the local K3 surface:
\begin{equation}\label{hilb}
\cK \simeq \IC^2 / \hat{D}_{g+1} \ , \qquad
\hat{D}_{g+1} = \gen{\mat{\omega_{2(g-1)} & 0 \\ 0 & \omega_{2(g-1)}^{-1}} \ , \mat{0 & i \\ i & 0}} \ , 
\end{equation}
with $\omega_n$ being the primitive $n$-th root of unity and $|\hat{D}_{g+1}| = 4(g-1)$.
The Hilbert series is given by
\begin{equation}
H(T) = \frac{1+T^{2g}}{(1-T^4)(1-T^{2(g-1)})} \ .
\end{equation}

\section{Modular Elliptic K3, Trivalent Graphs and $\cN=2$ Gauge Theory}
\setcounter{footnote}{0}
In the previous section, we have seen the emergence of particular trivalent graphs as Schreier-Cayley coset graphs and as skeleton diagrams for $\cN=2$ supersymmetric field theories in four dimensions with product $SU(2)$ gauge groups.
So too, have we seen the emergence of K3 surfaces, as modular elliptic surfaces and as the Kibble branch of the moduli space of the gauge theory.
It is therefore natural that we juxtapose these particulars in an intriguing correspondence.

As mentioned at the end of \sref{s:modSurface}, modular elliptic K3 surfaces arise for index 24 subgroups.
Specifically, according to \cite{mckaysebbar}, these congruence subgroups are, from the the list of the 33 presented in the previous section and quoted from cit.~ibid.:
$\Gamma(4)$, $\Gamma_0(3) \cap \Gamma(2)$, $\Gamma_1(7)$, $\Gamma_1(8)$, 
$\Gamma_0(8) \cap \Gamma(2)$, $\Gamma_0(12)$, $\Gamma_0(16)$, $\Gamma(8;4,1,2)$ and $\Gamma(16;16,2,2)$, a total of 9.
They organize themselves into 4 families: (I) $\Gamma(4)$ and $\Gamma(8; 4,1,2)$ are index 2 subgroups of $\Gamma_0(4) \cap \Gamma(2)$, itself an index 12 subgroup of the modular group; (II) $\Gamma_0(3) \cap \Gamma(2)$ and $\Gamma_0(12)$ are index 2 subgroups of $\Gamma_0(6)$; (III) $\Gamma_1(8)$, $\Gamma_0(8) \cap \Gamma(2)$, $\Gamma_0(16)$ and $\Gamma(16;16,2,2)$ are 4 index 2 subgroups of $\Gamma_0(8)$; (IV) $\Gamma_1(7)$.

\begin{table}[!h!t]
\[\hspace{-2cm}
\begin{array}{|c|c|c|c|c|}\hline
&\mbox{Group} & \mbox{Substitution} & \mbox{Parent Subgroup} & \mbox{Equation}
\\ \hline \hline
$Ia$ & \Gamma(4) & s \to (s^2-1) / (s^2+1) & \Gamma_0(4) \cap \Gamma(2) & 
x(x^2+2y+1)+s(x^2-y^2)) = 0 \\ \hline
$Ib$ & \Gamma(8; 4,1,2) & s \to s^2+1 & & \\ \hline \hline

$IIa$ & \Gamma_0(3) \cap \Gamma(2) & s \to 8s^2/(8-s^2) & \Gamma_0(6) &
(x+y)(x+1)(y+1)+sxy = 0 \\ \hline
$IIb$ & \Gamma_0(12) & s \to 1-s^2 & & \\ \hline \hline

$IIIa$ & \Gamma_1(8) & s \to s^2/(s^2+1) & \Gamma_0(8) &
(x+y)(xy-1)+ 4 i s x y = 0
\\ \hline
$IIIb$ & \Gamma_0(8) \cap \Gamma(2) & s \to (s^2-1)/(s^2+1) & & \\ \hline
$IIIc$ & \Gamma_0(16) & s \to s^2 & & \\ \hline
$IIId$ & \Gamma(16;16,2,2) & s \to s^2+1 & & \\ \hline  \hline

$IV$ & \Gamma_1(7) & - & - & 
\begin{array}{r}
y^2 + (1+s-s^2)xy + (s^2-s^3) y \\
= x^3 + (s^2-s^3)x^2
\end{array}
\\ \hline
\end{array}
\]
\caption{{\sf \small
The nine index 24, genus 0, torsion-free subgroups of the modular group and the explicit Weierstra\ss\ form of the defining equations of the associated elliptically fibred modular K3 surfaces. We have grouped the 9 subgroups according to some intermediate subgroup, marked ``parent subgroup'', of which each family is an index 2 subgroup thereof (signifying that the latter is actually an index 12 subgroup of the modular group). The defining equation of the modular surface of the parent is given to the right and that of the 9 subgroups of our concern can be obtained by the substitution rule on the parameter $s$ as indicated.
}}
\label{t:eq24}
\end{table}

\begin{figure}[h!t!]
\centerline{
\includegraphics[trim=0mm 0mm 0mm 0mm, clip, width=5.5in]{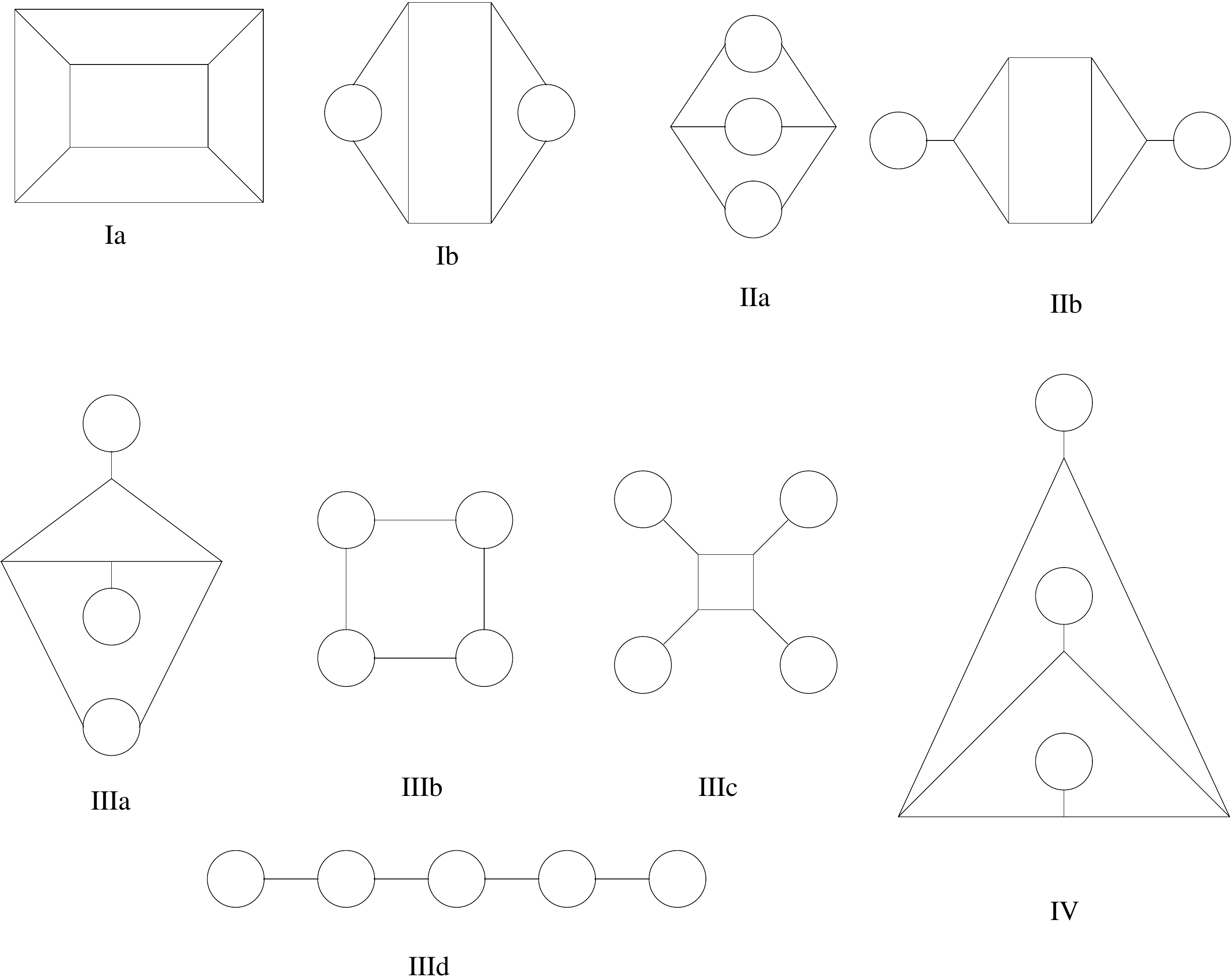}
}
\caption{{\sf {\small
The Schreier-Cayley coset graphs of the 9 index 24, genus 0, torsion-free congruence subgroups of the modular group. Specfically, they are:
(Ia) $\Gamma(4)$,
(Ib) $\Gamma(8; 4,1,2)$;
(IIa) $\Gamma_0(3) \cap \Gamma(2)$, (IIb) $\Gamma_0(12)$;
(IIIa) $\Gamma_1(8)$, (IIIb) $\Gamma_0(8) \cap \Gamma(2)$, (IIIc) $\Gamma_0(16)$  (IIId) $\Gamma(16;16,2,2)$; (IV) $\Gamma_1(7)$.
All graphs are trivalent with 12 edges, 8 nodes and 5 independent closed circuits (drawn on a sphere it has 6 faces).
}}
\label{f:graph24}}
\end{figure}

In \tref{t:eq24}, we collect the results of \cite{mckaysebbar} and \cite{TopYui} to tabulate the subgroups with the explicit equations of the associated elliptic K3 surfaces as algebraic equations in affine coordinates $(x,y)$, depending on the base projective coordinate $s$.
The equation can be obtained from the (intermediate) parent subgroup via a substitution of the base coordinate, as shown in the table.
Moreover, using the contracted-node notation discussed in \fref{f:coset-eg}, we present the Schreier-Cayley coset graphs of these 9 index 24 congruence subgroups in \fref{f:graph24} (cf.~\cite{mckaysebbar}).

\comment{
OLD RESULT-----------
\begin{array}{|c|c|}\hline
\Gamma(4) & y^2 = x(x-1)(x-\frac14 ( t + 1/t)^2) \\ \hline
\Gamma_0(12) & y^2 + (t^2+1)xy - t^2(t^2-1)y = x^3 -t^2(t^2-1)x^2 \\ \hline
\Gamma_0(16) & \\ \hline
\Gamma_0(3) \cap \Gamma(2) & 
(s^2 - 8)(x + y)(x + 1)(y + 1) = 8s^2xy \\ \hline
\Gamma_0(8) \cap \Gamma(2) & \\ \hline
\Gamma_1(7) & y^2 + (1+t-t^2)xy + (t^2-t^3) y = x^3 + (t^2-t^3)x \\ \hline
\Gamma_1(8) & \\ \hline
\Gamma(16;16,2,2) & \\ \hline
\Gamma(8; 4,1,2) & y^2 = 
 x^3 - 2(8t^4¡Ý16t^3+ 16t^2 - 8t+ 1)x^2 + (8t^2-8t+ 1)(2t-1)^4 x \\ \hline
\Gamma(4) & (s^2+1)x(x^2+2y+1)+(s^2-1)(x^2-y^2)=0 \\ \hline
\Gamma(8; 4,1,2) & x(x^2+2y+1)+(s^2+1 (x^2-y^2)=0 \\ \hline 
\hline
\Gamma_0(3) \cap \Gamma(2) & 
(s^2 - 8)(x + y)(x + 1)(y + 1) = 8s^2xy \\ \hline
\Gamma_0(12) & (x+y)(x+1)(y+1)+(1-s^2)xy = 0 \\ \hline
\hline
\Gamma_1(8) & \\ \hline
\Gamma_0(8) \cap \Gamma(2) & \\ \hline
\Gamma_0(16) & \\ \hline
\Gamma(16;16,2,2) & \\ \hline
\hline
\Gamma_1(7) & y^2 + (1+s-s^2)xy + (s^2-s^3) y = x^3 + (s^2-s^3)x \\ \hline
\end{array}
\end{equation}
}

\comment{
It is expedient to find the explicit change of coordinates to bring the two sets of K3 surfaces into each other.
More precisely, we look for a bi-rational transformation taking $\IC[x,y,s]$ to $\IC[u,v,w]$ from the forms in \tref{t:eq24} to that in \eqref{K3-D6}.
}

We can regard these coset graphs as skeleton diagrams of $\cN=2$ theories.
We note that they all have $(g,e) = (5,0)$; the number of independent closed circuits $g$ can be read off by noting that they all have no external lines and thus $8=2g-2$ nodes and $12=3g-3$ internal edges.
Indeed, we emphasize that $g$ of the diagram is the genus of the Riemann surface $\Sigma_G$ on which M5-branes can be wrapped to produce the gauge theory; this is different from the genus of the modular curve associated to the subgroup which gives the coset graph, which we recall to be 0 by construction.
Subsequently, by \eqref{hilb}, we have that the Kibble branch of the moduli space of all these $\cN=2$, $SU(2)^{12}$  gauge theories is isomorphic to the local K3 surface $\IC^2 / \hat{D}_6$, the standard defining equation for which is
\begin{equation}\label{K3-D6}
\{u^2 + v^2 w = w^5\} \subset \IC[u,v,w] \ .
\end{equation}

For the sake of our ensuing analyses, it is expedient to recast the affine equations in \eqref{t:eq24} into standard form, treating $s$ as a complex parameter.
First, let us recall some elementary facts on canonical forms of elliptic curves and surfaces.

\subsection{Standard Form and $J$-Invariant}
The standard {\em Tate form} of the elliptic curve (q.~v., e.~g.~\cite{serre,Milne}) is
\begin{equation}\label{tate}
y^2 + a_1 x y + a_3 y = x^3 + a_2 x^2 + a_4 x + a_6 \ ;
\end{equation}
upon defining 
\begin{equation}
b_2 = a_1^2+4a_2,\quad 
b_4 = a_1a_3+2a_4, \quad
b_6 = a_3^2+4a_6,\quad 
b_8 = a_1^2a_6-a_1a_3a_4+a_2a_3^2+4a_2a_6-a_4^2 \ ,
\end{equation}
the discriminant $\Delta$ and the $J$-invariant become
\begin{equation}
\Delta = -b_2^2b_8+9b_2b_4b_6-8b_4^3-27b_6^2 \ , \qquad
J = \frac{b_2^2 - 24 b_4}{\Delta} \ .
\end{equation}
Note that we are not including the famous 1728 factor here for convenience and for comparing with \cite{mckaysebbar}, thus, as is customary, we use an upper-case $J$, rather than the lower-case.

One could also decouple $x$ and $y$ and bring the Tate form, via the substitution $y \to y - \frac12(a_1 x + a_3)$, to
\begin{eqnarray}
\nn y^2 &=& x^3 + x^2 \left(\frac{a_1^2}{4}+a_2\right) + x
   \left(\frac{a_1 a_3}{2} + a_4\right)+\frac{a_3^2}{4}+a_6 \\
\label{factorE}
&=& (x-\lambda_1)(x-\lambda_2)(x-\lambda_3)
\end{eqnarray}
with $\lambda_{i=1,2,3}$ the three roots of the cubic in $x$.
In this case, the $J$-invariant\footnote{
Equivalently one could redefine $x$ to the right hand side of \eqref{factorE} and put it into Weierstra\ss\ form $y^2 = 4 x^3 + g_2 x + g_3$, whereupon
the $J$-invariant becomes $g_2^3 / (g_2^3 - 27 g_3^2)$.
} can be written as
\begin{equation}
J = \frac{256(1-\lambda+\lambda)^3}{\lambda^2(1-\lambda)^2} \ , \qquad
\lambda = \frac{\lambda_j - \lambda_k}{\lambda_j - \lambda_i} \ ,
\end{equation}
where $i,j,k$ are any distinct choices from $1,2,3$.
We can easily check that 6 possible combinations arise in this cross ratio 
for  $\lambda$, all of which give the same expression for $J$, as must be so.

We readily apply the above standard techniques to our list of explicit equations in \tref{t:eq24}.
The only subtlety is that we have elliptic surfaces written as fibrations over a base $\IP^1_C$ with coordinate $s$; this simply means that the Tate coeffcients $a_{1,2,3,4,6}$ will be rational functions of $s$ and thus reside in the function field $\IC(s)$. 
We summarize all cases of interest in \tref{t:weier}.
Note that we have done a little more and actually brought the equations from Tate form to a Weierstra\ss\ form where the $x$ and $y$ variables decouple so that the coefficients $a_1,a_3$ vanish.
For reference we have also computed all the $J$-invariants of the curves, also as functions of the base $\IP^1_C$-coordinate $s$.

\begin{table}[h!]
\[
\begin{array}{|c|c|l|}\hline
&\mbox{Group} & \mbox{Standard Form: } \{a_1,a_2,a_3,a_4,a_6\} 
\bigstrut[tb] \\ \hline \hline
$Ia$ & \Gamma(4) & 
\{
0 , \frac{8 \left(s^4+1\right)}{\left(s^2+1\right)^2} , 0 ,\frac{16
   \left(s^2-1\right)^2}{\left(s^2+1\right)^2},0
\}
\\[4mm]&&\qquad\qquad J=
\frac{16(1+14s^4+s^8)^3}{s^4(s^4-1)^4}
\bigstrut[tb]\\ \hline
$Ib$ & \Gamma(8; 4,1,2) & 
\{
0 , 4 \left(s^2+1\right)^2+4 , 0 , 16 \left(s^2+1\right)^2 , 0 
\}
\\[4mm]&&\qquad\qquad J=
\frac{256 \left(\left(s^2+2\right)
   \left(s^3+s\right)^2+1\right)^3}{\left(s^2+2\right)^2 \left(s^3+s\right)^4}
\bigstrut[tb] \\ \hline

$IIa$ & \Gamma_0(3) \cap \Gamma(2) & 
\{0,\frac{4 \left(3 s^2+8\right)^2}{\left(s^2-8\right)^2},0,\frac{128 s^2 \left(3
   s^2+8\right)}{\left(s^2-8\right)^2},\frac{1024 s^4}{\left(s^2-8\right)^2}
\}
\\[4mm]&&\qquad\qquad J= 
\frac{\left(3 s^2+8\right)^3 \left(3 s^6+600 s^4-960 s^2+512\right)^3}{8 s^6 \left(8-9
   s^2\right)^2 \left(s^2-8\right)^6}
\\ \hline
$IIb$ & \Gamma_0(12) & 
\{
0 , \left(s^2-3\right)^2 , 0 , 8 \left(s^4-4 s^2+3\right) , 16 \left(s^2-1\right)^2 
\}
\\[4mm]&&\qquad\qquad J=
\frac{\left(s^8-12 s^6+30 s^4-12 s^2+9\right)^3}{s^4 \left(s^2-9\right)
   \left(s^2-1\right)^3} 
\bigstrut[tb]\\ \hline

$IIIa$ & \Gamma_1(8) & 
\{
0,-\frac{16 s^4}{\left(s^2+1\right)^2}-4,0,\frac{128
   s^4}{\left(s^2+1\right)^2},-\frac{256 s^4}{\left(s^2+1\right)^2}
\}
\\[4mm]&&\qquad\qquad J=
-\frac{16 \left(s^8-28 s^6-10 s^4+4 s^2+1\right)^3}{s^4 \left(s^2+1\right)^8 \left(2
   s^2+1\right)}
\bigstrut[tb]\\ \hline
$IIIb$ & \Gamma_0(8) \cap \Gamma(2) & 
\{
0,-\frac{16 \left(s^2-1\right)^2}{\left(s^2+1\right)^2}-4,0,\frac{128
   \left(s^2-1\right)^2}{\left(s^2+1\right)^2},-\frac{256
   \left(s^2-1\right)^2}{\left(s^2+1\right)^2}
\}
\\[4mm]&&\qquad\qquad J=
-\frac{4 \left(s^8-60 s^6+134 s^4-60 s^2+1\right)^3}{s^2 \left(s^2-1\right)^2
   \left(s^2+1\right)^8}
\bigstrut[tb]\\ \hline
$IIIc$ & \Gamma_0(16) & 
\{
0,-4 \left(4 s^4+1\right),0,128 s^4,-256 s^4
\}
\\[4mm]&&\qquad\qquad J=
\frac{16 \left(16 s^8-16 s^4+1\right)^3}{s^4 \left(s^4-1\right)}
\bigstrut[tb]\\ \hline
$IIId$ & \Gamma(16;16,2,2) &  
\{
0,-16 \left(s^2+1\right)^2-4,0,128 \left(s^2+1\right)^2,-256
   \left(s^2+1\right)^2
\}
\\[4mm]&&\qquad\qquad J=
\frac{16 \left(16 s^8+64 s^6+80 s^4+32 s^2+1\right)^3}{\left(s^2+2\right)
   \left(s^3+s\right)^2}
\bigstrut[tb]\\ \hline

$IV$ & \Gamma_1(7) & 
\{
 0 , \left(-s^2+s+1\right)^2 , 0 , 8 (s-1) s^2 ((s-1) s-3) , 16 (s-1)^2 s^4
\}
\\[4mm]&&\qquad\qquad J=
\frac{\left(s^8-12 s^7+42 s^6-56 s^5+35 s^4-14 s^2+4 s+1\right)^3}{(s-1)^7 s^7
   \left(s^3-8 s^2+5 s+1\right)}
\bigstrut[tb]\\ \hline
\end{array}
\]
\caption{{\sf \small
The explicit equations, in Tate-Weierstra\ss\ form, for the elliptic K3 surfaces corresponding to the nine index 24, genus 0, torsion-free subgroups of the modular group.
We have given the standard coefficients of the equation, $\{a_1,a_2,a_3,a_4,a_6\}$, as explained in \eqref{tate}.
The variable $s$ is the coordinate on the base $\IP^1_C$ over which $x,y$ are affine coordinates of the elliptic fibration.
The $J$-invariants of the curves, as functions of $s$, are also recorded.
}}
\label{t:weier}
\end{table}

The $J$-invariants given in \tref{t:weier} can be compared with the more complicated forms, obtained via other means, presented in \cite{mckaysebbar}.
Indeed, one of the chief purposes of \cite{TopYui} was to find simpler expressions for the explicit elliptic surfaces through the method of coverings.
Of course, we could apply a $GL(2;\IC)$ transformation on $s$, amounting to a mere change of the base $\IP^1_C$ coordinates, to $j(s)$ in the table, and obtain the forms of $j(t)$ in \cite{mckaysebbar}.
For example, for group Ia, the change $s \mapsto t/2$, gives that of the corresponding result tabulated in \S3 of \cite{mckaysebbar}.
Similarly, for group Ib, the more involved transformation $s \mapsto \frac{i \sqrt{2} (t-1)}{t-3}$ suffices.

One could also read off, from the dependence on $s$ of the Tate coefficients $a_i$ in the defining equations, the exact type of singular fibres of the various elliptic surfaces in the Kodaira classification.
Now, our elliptic surfaces are semi-stable \cite{mckaysebbar} thus in particular all our index 24 K3 surfaces have singular fibres of type $I_n$ where $n$ is the cusp width, the tabulation of which is given in \S3 of \cite{mckaysebbar}; we will return to the cusp widths in the next subsection.
As a comparison, the affine dihedral singularity for our gauge theory given in \eqref{K3-D6} is type $I^*_{2}$.
It is interesting, however, that the index 6 subgroup $\Gamma(2)$, which we have mentioned several times, gives exactly this $\hat{D}_6$ singularity.

\subsection{Extremality, Belyi Maps and Dessins d'Enfant}\label{s:dessin}
One thing special about our nine elliptic K3 surfaces is that they are {\it extremal} \cite{sebbar,TopYui,shioda-rev} in the sense of having maximal rank of the Picard group\footnote{While at the same time having finite Mordell-Weil group.
We will return to this point later in \sref{s:arithmetic}.}.
A classification of such surfaces is given in \cite{MP}.
A remarkable fact about these extremal elliptic surfaces, which are quite rare, is that their $j$-functions, treated as a rational map from the base $\IP^1_C$ to $\IP^1$, are ramified at only $(0,1,\infty)$ in the target \cite{shioda-rev}.
In other words, and in our normalization, this means that
\begin{equation}
s \in \IP^1_C \mapsto J(s) \in \IP^1 : 
\frac{d}{d s} \left. J(s)\right|_{\tilde{s}} = 0 \Leftrightarrow 
  J(\tilde{s}) = 0, \ 1728, \mbox{ or } \infty \ ,
\end{equation}
for our list of $J$-invariants in \tref{t:weier}.

Maps to $\IP^1$ ramified only at $(0,1,\infty)$ are called {\it Belyi} maps and can be represented graphically as Grothendieck's {\it dessins d'enfant}.
Remarkably, this structure has appeared in the context of $\cN=2$ gauge theories in four dimensions in \cite{hep-th/0611082} and we will turn to this shortly.
Recently, the dessins have also been crucial \cite{Jejjala:2010vb,Hanany:2011ra} to the understanding of so-called {\it dimer models} or {\it brane tilings} \cite{Hanany:2005ve,Feng:2005gw} in the context of world-volume gauge theories on D-branes probing non-compact toric Calabi-Yau manifolds.

Let us first collect data and compute the branch and ramification points as well as the ramification indices for the $J$-maps in \tref{t:weier}.
Take group Ia, $\Gamma(4)$.
Here, $J = {16(1+14s^4+s^8)^3}{s(s^4-1)}^{-4}$.
We obtain the branch points by solving $\frac{d}{ds} \left. J(s) \right|_{\tilde{s}} = 0$.
We find that at 8 points
\begin{equation}
\tilde{s} = \left( \pm \frac12 \pm \frac{i}{2} \right) \left( \pm 1 \pm \sqrt{3} \right) \leadsto J(\tilde{s}) = 0 \ ;
\end{equation}
moreover, at all these points, the series expansion of $J(s)$ vanishes up to order 3, i.e., $J(s) \sim C (s-\tilde{s})^3 + \cO((s-\tilde{s})^4)$ for constant $C$. Hence, the ramification index at all these 8 pre-images of 0 is 3.

Next, we have that at 12 points
\begin{equation}
\tilde{s} = \{
\exp(\frac{\pi i n}{4}) \ , \pm 1 \pm \sqrt{2} \ ,
\pm i (1 + \sqrt{2}) \ , \pm i (-1 + \sqrt{2})
\} \leadsto J(\tilde{s}) = 1728 \ ,
\end{equation}
where $n=0,1,2,3$.
At all these points, the series expansion of $J(s)$ goes as $1728 + C (s-\tilde{s})^2 + \cO((s-\tilde{s})^3)$ for numerical $C$.
Hence, the ramification index is 2 for all 12 pre-images of 1728.

Finally, we have that at the 6 points
\begin{equation}
\tilde{s} = \{0 \ , \pm 1 \ , \pm i \ , \infty \} \leadsto
J(\tilde{s}) = \infty \ .
\end{equation}
At the 5 finite points, the series expansion of $1/J(s)$ proceeds as 
$\sim C (s-\tilde{s})^4 + \cO((s-\tilde{s})^5)$, so that the ramification index at these 5 pre-images of $\infty$ is 4.
At $\infty$, we clearly have that $J(s) \sim s^{8 \cdot 3 - 4 - 4\cdot 4}$, so that the ramification index is also 4.

Though in this case all ramification indices are equal for each pre-image, the situation in general may not be so.
In fact, as we shall soon see, for the purposes of constructing the dessins, it suffices to know the ramification indices only and the exact values of the pre-images, which often assume extremely complicated expressions, need not trouble us.
Therefore, let us adopt the convenient notation from \cite{Hanany:2011ra}: 
let there be $W$ pre-images of $0$, $B$ pre-images of $1728$ and $I$ pre-images of $\infty$.
Furthermore, let the ramifications of the pre-images of these three marked points be, respectively,
$\{r_0(1) , r_0(2) , \ldots , r_0(B)\}$,
$\{r_1(1) , r_1(2) , \ldots , r_1(W)\}$, and
$\{r_\infty(1) , r_\infty(2) , \ldots , r_\infty(I)\}$.
Then, the structure of the $J(s)$-map as a Belyi map can be recorded by the triple
\begin{equation}
J(s) \sim \left\{
\begin{array}{l}
r_0(1) , r_0(2) , \ldots , r_0(B) \\
r_1(1) , r_1(2) , \ldots , r_1(W) \\
r_\infty(1) , r_\infty(2) , \ldots , r_\infty(I)
\end{array}
\right\} \ .
\end{equation}
Hence, for the above example, we can write ${\scriptsize \left\{ \begin{array}{l}
3, 3, \ldots, 3 \\
2, 2, 2, \ldots, 2 \\
4, \ldots 4
\end{array}\right\}}$, or 
${\scriptsize \left\{  \begin{array}{l}
3^8 \\
2^{12} \\
4^6
\end{array}\right\}}$ for short.
It is straightforward to
compute all the ramification indices for the nine $J(s)$-maps and collect the results in \tref{t:ram}.

\begin{table}
\[
\begin{array}{|l|l|}\hline
\hline
\mbox{Group} & \mbox{Ramification} \\ \hline \hline
$Ia$ : \Gamma(4) & 
{\scriptsize 
\left\{  \begin{array}{l}
3^8 \\
2^{12} \\
4^6
\end{array}\right\}}
\\ \hline
$Ib$ : \Gamma(8; 4,1,2) & 
{\scriptsize 
\left\{  \begin{array}{l}
3^8 \\
2^{12} \\
2^2, 4^3, 8
\end{array}\right\}}
\\ \hline
$IIa$ : \Gamma_0(3) \cap \Gamma(2) & 
{\scriptsize 
\left\{  \begin{array}{l}
3^8 \\
2^{12} \\
2^3, 6^3
\end{array}\right\}}
\\ \hline
$IIb$ : \Gamma_0(12) &
{\scriptsize 
\left\{  \begin{array}{l}
3^8 \\
2^{12} \\
1^2, 3^2, 4, 12
\end{array}\right\}}
\\ \hline
\end{array}
\quad \
\begin{array}{|l|l|}\hline
\hline
\mbox{Group} & \mbox{Ramification} \\ \hline \hline
$IIIa$ :  \Gamma_1(8) & 
{\scriptsize 
\left\{  \begin{array}{l}
3^8 \\
2^{12} \\
1^2, 2, 4, 8^2
\end{array}\right\}}
\\ \hline
$IIIb$ : \Gamma_0(8) \cap \Gamma(2) & 
{\scriptsize 
\left\{  \begin{array}{l}
3^8 \\
2^{12} \\
2^4, 8^2
\end{array}\right\}}
\\ \hline
$IIIc$ : \Gamma_0(16) & 
{\scriptsize 
\left\{  \begin{array}{l}
3^8 \\
2^{12} \\
1^4, 4, 16
\end{array}\right\}}
\\ \hline
$IIId$ : \Gamma(16;16,2,2) &
{\scriptsize 
\left\{  \begin{array}{l}
3^8 \\
2^{12} \\
1^2, 2^3, 16
\end{array}\right\}}
\\ \hline
$IV$ : \Gamma_1(7) & 
{\scriptsize 
\left\{  \begin{array}{l}
3^8 \\
2^{12} \\
1^3, 7^3
\end{array}\right\}}
\\ \hline
\end{array}
\]
\caption{\sf {\small
The ramification indices, at the various pre-images of 0, 1728 and $\infty$, of the $J$-invariant of our 9 modular elliptic K3 surfaces, treated as Belyi maps from the base $\IP^1_C$ of the elliptic fibration onto a target $\IP^1$.
}}
\label{t:ram}
\end{table}

To draw the dessin is simple:
one marks one black node for the $i$-th pre-image of 0, with $r_0(i)$ edges emanating therefrom, similarly, one marks one white node for the $j$-th pre-image of 1 (here normalized to 1728), with $r_1(j)$ edges.
Thus we have a bipartite graph with $B$ black nodes and $W$ white nodes.
Now we connect the nodes with the edges - joining only black with white - such that $I$ faces, each being a polygon with $2r_{\infty}(k)$ sides.
In order to fully specify the connectivity, we need to know, in addition, the possible monodromies around the ramification point.
Again, this is nicely catalogued in \cite{MP} and the explicit equations are discussed in \cite{BM}.

We note that in all our 9 cases, the ramifications over 0 are of index 2, those of 1728 are of index 3, as is shown to be so required in \cite{MP,BM}.
Returning to our \tref{t:weier}, in addition to comparing to the results obtained in \cite{LY} from the Tate algorithm, we should once more consult the complete classification of the 112 extremal semi-stable elliptic K3 surfaces with exactly 6 singular fibres of which our 9 cases is a special subset.
In their notation, the surfaces are determined by the ramification indices at $\infty$, which are referred to as {\em cusp widths}.
Therefore, the bottom entry to our ramification indices in \tref{t:ram} is precisely the last column on ``Widths'' in the table in Section 3 of \cite{mckaysebbar}.

\subsection{Arithmetic}\label{s:arithmetic}
Having discussed extremality and dessins d'enfant, it is hardly resistible to venture into the arithmetic properties of our K3 surfaces.
Indeed, the number-theoretical properties of moduli spaces of vacua for supersymmetric gauge theories was recently discussed in \cite{He:2010mh}.
It is therefore fruitful for us to briefly discuss our theories under this light.

A general lesson about the arithmetic of Calabi-Yau manifolds roughly goes as follows: a Calabi-Yau $n$-fold, under special restrictions, should be associated with a modular form of weight $n+1$.
Indeed, the case of $n=1$ is the celebrated result of Shimura-Taniyama-Wiles, that every elliptic curve has an $L$-series which is a cusp form of weight 2 and level equal to the conductor of the curve.
The situation in dimension $n>1$ becomes much more restrictive, and the reader is referred to \cite{Schuett,Meyer} for some recent progress.

For K3 surfaces, for which $n=2$, a nice theorem of Livn\'e \cite{Livne} states that singular K3 surfaces are modular in that their L-functions are newforms of weight 3.
Luckily this restriction of ``singular'' is precisely the case at hand.
We recall that a ``singular'' K3 in the literature has come to refer to its
{\it exceptionality} rather than geometric singularity; it is a K3 surface with
maximal Picard number $\rho = 20$.
Our extremal elliptic K3 surfaces are indeed prime examples of such singular K3 surfaces \cite{shioda-rev}.
In fact, as mentioned before, the extremality further requires that our surfaces have finite Mordell-Weil groups.
These have been computed in \cite{shimada} (presented therein as a part of their extensive list of all extremal elliptic K3 surfaces); for reference we include them here:
\begin{equation}
\begin{array}{|l|l|}\hline
\hline
\mbox{Group} & \mbox{Mordell-Weil} \\ \hline \hline
$Ia$ : \Gamma(4) & 
\IZ_4 \times \IZ_4 
\\ \hline
$Ib$ : \Gamma(8; 4,1,2) & 
\IZ_2 \times \IZ_4 
\\ \hline
$IIa$ : \Gamma_0(3) \cap \Gamma(2) & 
\IZ_2 \times \IZ_6 
\\ \hline
$IIb$ : \Gamma_0(12) &
\IZ_3 \mbox{ or } \IZ_6 
\\ \hline
\end{array}
\quad \
\begin{array}{|l|l|}\hline
\hline
\mbox{Group} & \mbox{Mordell-Weil} \\ \hline \hline
$IIIa$ :  \Gamma_1(8) & 
\IZ_8  
\\ \hline
$IIIb$ : \Gamma_0(8) \cap \Gamma(2) & 
\IZ_2 \times \IZ_4 
\\ \hline
$IIIc$ : \Gamma_0(16) & 
\IZ_3 
\\ \hline
$IIId$ : \Gamma(16;16,2,2) &
\IZ_3 
\\ \hline
$IV$ : \Gamma_1(7) & 
\IZ_7 
\\ \hline
\end{array}
\end{equation}

\comment{
The arithmetic properties of our gauge theory moduli space, on the other hand, is quite simple.
Recalling the explicit equation of the $\hat{D}_6$ K3 surface from \eqref{K3-D6}, we see that we can re-write it as $u^2 = w(w^2+v)(w^2-v)$
}


\subsection{Seiberg-Witten Curves and Dessins}\label{s:SW-dessin}
Let us now return to gauge theories.
One of the most important quantities in supersymmetric gauge theories, especially those with $\cN=2$ supersymmetry, is the Seiberg-Witten curve.
This is a hyper-elliptic curve whose periods completely specify the spectra, coupling and low-energy effective Lagrangian, as well as non-perturbative information of the gauge theory.
In \cite{hep-th/0611082}, a unique perspective was taken on the Seiberg-Witten curve from the dessin point of view, of which we will now briefly remind the reader before relating it to our results in \sref{s:dessin}.

For $\cN=2$ pure $U(N)$ gauge theory, the Seiberg-Witten curve is the hyper-elliptic curve
\begin{equation}
\{y^2 = P_N(x)^2 - 4 \Lambda^{2 N} \} \subset \IC[x,y] \ ,
\end{equation}
where $P_N(x)$ is a degree $N$ polynomial in $z$ with coefficients depending explicitly on mass and coupling parameters and $\Lambda$ is some cut-off scale, which we can customarily take to be unity.
The factorization of the right hand side $P_N(x)^2 - 4 \Lambda^{2N}$, which in general we will call the non-trivial portion of the Seiberg-Witten curve, governs the position in moduli space where monopoles condense to give us confinement.
The map $\beta(x) = 1 - \frac{P_N(x)^2}{4 \Lambda^{2N}}$ was shown in \S2.6 of \cite{hep-th/0611082} to be a Belyi map at isolated points in the moduli space.
In general, one can construct (cf.~Eqs (2.5) and (2.6) in cit.~ibid.) polynomials $A(x)$ and $B(x)$ such that
\begin{equation}
A(x) - B(x) = P_N(x)^2 \ , \qquad
\beta(x) = 1 + \frac{P_N(x)^2}{B(x)} = \frac{A(x)}{B(x)} \ ,
\end{equation}
where $\beta(x)$ is a Belyi map and $P_N(x)$ is monic of degree $N$ and to be interpreted as the right side of the Seiberg-Witten curve of an $\cN=2$ gauge theory.
Moreover, $\beta$ is {\it clean} Belyi in the sense that all ramification indices at 1 are equal to 2.

\begin{figure}[!h!t!b]
\centerline{
\includegraphics[trim=0mm 0mm 0mm 0mm, clip, width=5.5in]{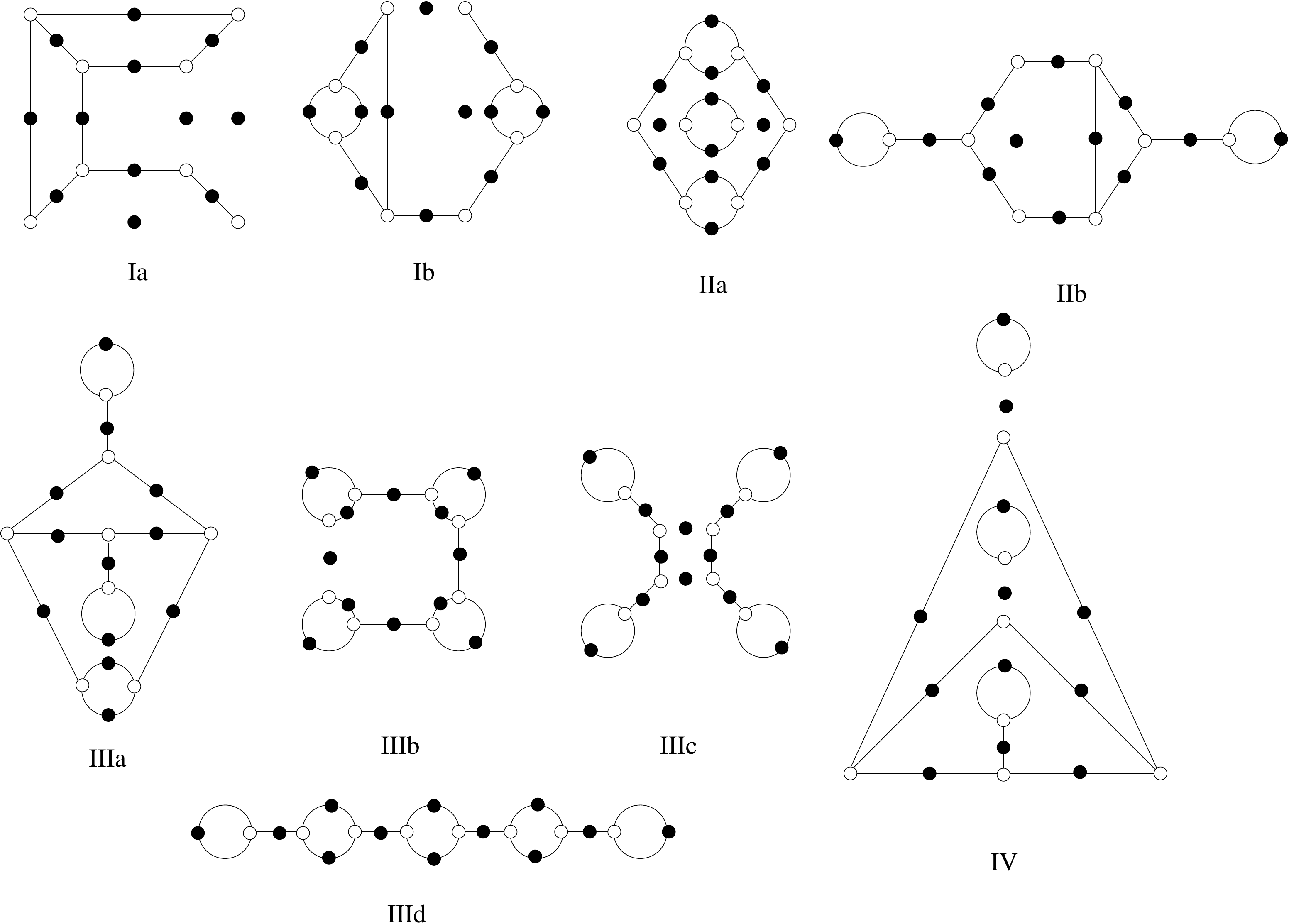}
}
\caption{{\sf {\small
The {\it dessins d'enfant} for the nine index 24, genus 0, torsion-free congruence subgroups of the modular group, drawn from the $J$-map of these groups when expressed as modular elliptic K3 surfaces fibred over $\IP^1_C$.
The $J$-maps are all clean Belyi rational maps.
}}
\label{f:dessin}}
\end{figure}

Remarkably, looking back at our list in \tref{t:ram}, the $J$-maps for our elliptic K3 surfaces are all {\it clean} Belyi: the ramification at all 12 pre-images of 1728 is exactly 2.
For reference, let us draw the dessins for our 9 players in \fref{f:dessin}.
Luckily, this is already done in the web-link in \cite{BM} and we largely reproduce them here, though we will mark the black/white nodes explicitly as an emphasis of the bipartite nature of the graphs.
Of course, we recognize the dessins in \fref{f:dessin} to be those in \fref{f:graph24}, but now decorated with bipartite black-white nodes.
The trivalency of the 8 nodes in  each graph signifies ramification index 3 at the 8 pre-images of 0, each of the 12 edges now has a black node inserted so as to encode the 12 pre-images of 1728, of ramification index 2, whereby making the Belyi map clean.

Given the explicit forms of our $J$-invariants as Belyi maps, it is easy to reverse engineer the right-hand side of an associated Seiberg-Witten curve.
We first take $J$ from \tref{t:weier}, normalize it by 1728 (to ensure the cleanliness to be at 1), and take the difference of the numerator and denominator:
\begin{equation}\label{PnJ}
P_{12}(x)^2 = \mbox{Numerator}(J(x)) - 1728 \mbox{ Denominator}(J(x)) \ .
\end{equation}
We make this computation and note that in the combination $\mbox{Numerator}(J(x)) - 1728 \mbox{ Denominator}(J(x))$ factorizes into perfect squares, whereby making the identification \eqref{PnJ} legitimate so that $P_{12}(x)$ is polynomial of degree 12.
This is a non-trivial factorization since all the numerators are perfect cubes and the denominators have factors of odd powers; such is, of course, a direct consequence of the clean nature of the dessin since we are finding the pre-images of 1 (or 1728 in our normalization) which all have ramification index 2.

For our present set of trivalent theories, we have $SU(2)^{12}$ with 8 tri-fundamental fields, obtainable by wrapping M5-branes over punctured Riemann surfaces whose am{\oe}ba projection \cite{Feng:2005gw} are the skeleton diagrams.
These punctured Riemann surfaces, denoted $\Sigma_G$ in \sref{s:phys}, have come to be known as {\it Gaiotto curves} and the Seiberg-Witten curves $\Sigma_{SW}$ for these generalized $SU(2)$ quiver theories were written down in \cite{Gaiotto:2009we}; they are double covers over the Gaiotto curves (in general for $N$ M5-brnaes, it is an $N$-fold cover).
More precisely, $\Sigma_{SW}$ are double covers over $\Sigma_G$ with non-trivial ramification at the punctured points and possibly other points as well which encode deformations in the Lagrangian.
The Seiberg-Witten curve is itself a hyper-elliptic curve of the form $y^2 = \phi(x)$ with $\phi(x) dx^2$ a quadratic differential having double poles at the punctured points and simple poles at the deformation points.

Our nine graphs all have $g=5$ and no external lines for global flavour symmetries, i.e., no punctures.
Hence, the Gaiotto curves are compact hyper-elliptic curves of genus 5.
To determine the genus of the Seiberg-Witten curve, we use the Riemann-Hurwitz formula that
\begin{equation}
2g(\Sigma_{SW}) - 2 = \deg(f) (2 g(\Sigma_G)-2) + R \ ,
\end{equation}
where $f$ is the double covering map, which thus is here of degree 2 and $R$ is the total degree of the ramification points.
Presently, $g(\Sigma_G) = 5$ and $R$ must come from points other than punctures, which, if we chose to be 4 points with simple poles of the quadratic differential, gives us $g(\Sigma_{SW}) = 11$.

Now, a hyper-elliptic curve of genus $g$ can be written as $y^2 = F_{2g + 2}(x)$ where $F_{2g + 2}(x)$ is a polynomial of degree $2g+2$ in $x$. 
Hence, the Seiberg-Witten curves for our 9 theories are of the form $y^2 = F_{24}(x)$.
In other words, the form of $P_{12}(x)$ in \eqref{PnJ} can prescribe the required Seiberg-Witten curves, since we have degree 24.
We remark here that there is subtlety as to whether this degree 24 hyper-elliptic curve is the Seiberg-Witten curve for an $U(12)$ (or, indeed, an $SU(12)$ theory as the $U(1)$ decouples in the low-energy limit), or that of the $SU(2)^{12}$ theory. The former is of rank 11, and is naturally associated with the genus 11 curve we just constructed; the latter is of rank 12. It would be interesting indeed to study any duality relations between these two theories in the present context\footnote{We thank Amihay Hanany and Diego Rodriguez-Gomez for pointing this out to us.
}.

Because of the rigidity of the $J$-maps, we are at special points in complex structure moduli space of $\Sigma_{SW}$.
This is reflected by the fact that we have no flavours and hence no mass parameters. Furthermore, the values of the couplings are fixed by the forms of the $J$-maps.
The only degree of freedom, in each case, is the cut-off scale $\Lambda$, at isolated values of which the right hand side $P_{12}^2(x) - 4 \Lambda^2=(P_{12}(x)+2\Lambda)(P_{12}(x)-2\Lambda)$ further factorizes to have multiple roots.
For example, for group Ia, at $\Lambda=2$, we have $y^2 = 16 s^4 \left(s^8-33 s^4-33\right) \left(s^{12}-33 s^8-33 s^4+2\right)$ so that two of the $U(1)$ factors condense and we are left with an unbroken $U(1)^{10}$ gauge group in the Coulomb branch.

Thus we have come full circle in our story here: 
\begin{enumerate}
\item We commenced with the generalized quivers of nine distinguished $\cN=2$ gauge theories in four dimensions which can be obtained by wrapping 2 M5-branes over a Gaiotto curve $\Sigma_G$; they all have K3 surfaces as the Kibble branch of vacua;
\item We re-interpreted them as the Schreier-Cayley coset graphs of 9 index 24 subgroups of the modular group, each of which gives a genus zero Riemann surface after factoring the upper half plane;
\item We then found the modular elliptic surfaces which are elliptic fibrations over these 9 torsion-free Riemann spheres (i,e., the base is a $\IP^1_C$); they all turn out to be K3 surfaces;
\item Subsequently, we computed the explicit $j$-invariants of these K3 surfaces as elliptic curves with a parameter being the coordinate $s$ of the base $\IP^1_C$;
\item We considered the $j$-invariants to be ramified maps to an auxiliary $\IP^1$.
These maps turn out to be clean Belyi and hence can be understood as the non-trivial portion of a Seiberg-Witten curve as a hyper-elliptic curve. 
\item Finally, these Seiberg-Witten curves are double covers of the Gaiotto curves with which we started, at special points in moduli space.
\end{enumerate}

\subsubsection{Dimer Models and $\cN=1$ Theories}
As a parting comment, we could also think of \tref{t:ram} as a list of $\cN=1$ four-dimensional quiver gauge theories using the rules prescribed in \cite{Hanany:2011ra}.
For example, the $\Gamma(4)$ entry can be read as a theory with 5 gauge group factors, 12 bifundamental or adjoint fields and 8 cubic terms in the superpotential.
It would be an gratifying future exercise to explicitly construct all these gauge theories according to the dessins and exploit their physical quantities, such as the classical moduli space of vacua.
\comment{
In this sense, we have gone through an intricate chain of dualities: we start with a trivalent graph, which simultaneous corresponds to a modular elliptic K3 surface as well as an $\cN=2$ generalized quiver gauge theory in four dimensions, we then compute the $J$-invariant of the Weierstra\ss\ equation for the K3 surfaces fibred over $\IP^1_C$ and recognize this to be a Belyi map, finally we use the Belyi map and construct, using the dimer-tiling-dessin technique, an auxilliary four-dimensional quiver gauge theory.
}

Of course, as emphasized above, the ramification indices are not enough to fully identify the Lagrangian of the gauge theory since we need to know permutation indices prescribed by the monodromies at the ramification points; luckily the dessins in \cite{BM} should give us all the possibilities.
Moreover, the gauge theories constructible from the dessins are traditionally from dimers drawn periodically on a plane, i.e., the pre-image of $\IP^1$ should be on a $T^2$. The above dessins are maps from $\IP^1_C$.
It is possible we could resort to the ``untwisting map'' introduced in \cite{Feng:2005gw} which change the pre-image torus to Riemann surfaces of other genera. 

One further subtlety is that we have ramification 1 at $\infty$ from the table, this means that we have isolated nodes in the dimer with 2-gons (closed circuit of 2 edges) attaching them to the rest. Therefore, in the gauge theory, we need to be careful to integrate out these massive modes.
These issues leave us with tantalizing future directions.

\subsubsection{Duality Groups}
One of the achievements of \cite{Gaiotto:2009we} is to generalize S-duality and Argyres-Seiberg duality for the large class of gauge theories given by the skeleton diagrams.
Indeed, theories with the same $(g,e)$ diagrammatically (which we recall are those with $e$ external legs and $g$ loops) are related to each other by S-duality, as is supported by the fact that they have the same moduli space of vacua, seen in our case in \eqref{K3-D6}.
The Hilbert series for the vacua were computed in \cite{Hanany:2010qu} for many classes of skeleton theories.

\comment{
More precisely,
the forms of the Seiberg-Witten curves in the present context of the trivalent gauge theories were also conjectured in \cite{Gaiotto:2009we}.
For $(g,e)$, these turn out to be Riemann surfaces of genus $g$ and $e$ marked points.
Hence, for our 9 examples, they are genus 5 hyper-elliptic curves.
We now proceed to discuss the duality structure of our theories as well as interpretations of the Seiberg-Witten curves.
}

It is first curious to note that some congruence subgroups have emerged in the seminal work on general $\cN=2$ gauge theories \cite{Seiberg:1994rs}. One can use the lifting technique of \cite{Witten:1997sc} to show that the Riemann surface  $\Sigma_G$ on which the M5-branes wrap have the monodromy matrices at certain singularities, and thus the S-duality group, generate precisely congruence subgroups of $PSL(2;\IZ)$.

In the case of $\Sigma_G$ having degree $n+1$ in the variable $t$ within the defining polynomial $F$, the S-duality group is $\pi_1(\cM_{0;n+3,2})$, the fundamental group of the moduli space of genus zero Riemann surfaces with $n+3$ distinct marked points, two of which are distinguished and ordered.
At $n=1$, for the simplest example, which corresponds to $\cN=2$, $SU(k)$ gauge theory in four dimensions, the duality group is $\pi_1(\cM_{0;4,2})$, which is known to be $\Gamma_0(2)$; it is most probably a coincidence that this happens to be a modular group.

In our present situation, the S-duality group for the generalized $\cN=2$ quiver theories with $g$ closed circuits and $e$ external lines was demonstrated in \cite{Gaiotto:2009we} to be $\pi_1(\cM_{g;e})$, with $\cM_{g;e}$ being the moduli space of genus $g$ Riemann surfaces with $e$ marked points.
Returning to \fref{f:graph24}, wherein all our trivalent graphs have $g=5$ and $e=0$, the duality group for the theories corresponding to interpreting the graphs as generalized $\cN=2$ quivers is thus $\pi_1(\cM_{5,0})$, habitually simply denoted as $\cM_5$.
It is well-known that the moduli space of genus $g$ Riemann surfaces is the quotient by the Teichm\"uller space, of complex dimension $3g-3$, by the mapping class group.
A classic result of \cite{patterson} showed that $\pi_1(\cM_{g,e})$ is trivial for $g > 2$ and is equal to $\IZ_5$ for $g=2$ and $e \equiv 4 \mod 5$ and so
our $g=5$ theories have trivial duality group.

\subsection{Seiberg-Witten Curves Revisited}\label{s:SW-revisit}
It is perhaps amusing, given the forms of the defining equations of our elliptic modular K3 surfaces, to consider them, especially in our current train of thought of gauge theories, as Seiberg-Witten curves themselves.
Interestingly, a similar vein of ideas had been taken in \cite{malmendier}.
There, the Seiberg-Witten curves for the canonical examples of supersymmetric gauge theories, viz., $\cN=2$ super-Yang-Mills with $SU(2)$ gauge group with $N_f$ massless hypermultiplets are treated as elliptic curves fibered over $\IP^1$ defined by the $u$-plane.

The observation of \cite{malmendier} is that for $N_f=2$, the Seiberg-Witten curve 
\begin{equation}
y^2 = 4x^3 - g_2 x - g_3 \ ,
\end{equation} 
with $g_2 = \frac13 u^2 + 1, \
g_3 = \frac{1}{27}u(u^2-9)$, can be thought of as a rational elliptic surface in $\IC[x,y,u]$.
In fact, it is a modular elliptic surface over the base curve $\Gamma(2) \backslash \cH$. Similarly, for $N_f=0$ (i.e., pure $SU(2)$), the Seiberg-Witten curve as above, but with $g_2 = \frac13 u^2 - \frac14, \ g_3 = \frac{1}{126}u(8u^2-9)$, is a rational elliptic surface and is a modular surface over $\Gamma_0(4) \backslash \cH$.
We note that $\Gamma(2)$ and $\Gamma_0(4)$ are the only two genus zero subgroups of index 6.
The reader is also referred to \cite{Tai:2010im} which uses the relations of \cite{Alday:2009aq} for trialities at such special points.

In general, for $N_f < 4$ flavours of fundamental matter (it is customary to take $N_f < 2N_c$ and here we also normalize the cut-off scale $\Lambda$ to unity), the $SU(2)$ theory has the following Seiberg-Witten curves, which we collect from the nice review of \cite{Eguchi:1997pu}:
\begin{eqnarray}
\nn N_f = 0 & : & y^2 = (x^2-1)(x-u) ; \\
\nn N_f = 1 & : & y^2 = x^2(x-u) + \frac14 m x - \frac{1}{64}; \\
\nn N_f = 2 & : & y^2 = (x^2-\frac{1}{64})(x-u) + \frac14 m_1m_2x 
    - \frac{1}{64}(m_1^2 + m_2^2); \\
\nn N_f = 3 & : & y^2 = x^2(x-u) -\frac{1}{64}(x-u)^2 + \frac14 m_1m_2m_3x  
    - \frac{1}{64} (m_1^2 + m_2^2 + m_3^2) (x-u) \\ \label{SW-su2}
    && \qquad  - \frac{1}{64}(m_1^2m_2^2 + m_1^2m_3^2 + m_2^2m_3^2) \ .
\end{eqnarray}
We have left the mass parameters $m_{1,2,3}$ of the flavours for completeness.

\paragraph{Seiberg-Witten Curves as Modular Elliptic Surfaces: }
It would therefore be an interesting exercise to map the explicit equations of our modular elliptic curves in \tref{t:weier} to the Seiberg-Witten curves of the above set of gauge theories.

Our strategy is thus clear: we could, where possible, directly observe the forms of the explicit Weierstrass equations, treating $s$ as a parameter, and compare with the various Seiberg curves in \eqref{SW-su2}, with $u$ and the masses as parameters. 
Alternatively, we could find the $J$-invariants of the various Seiberg curves and see, by a possible linear fractional transformation, whether they correspond to any in the list in \tref{t:weier}.

Note that this is an entirely different approach from that undertaken in \sref{s:SW-dessin} because we are using the elliptic surfaces directly and viewing them as Seiberg-Witten curves of some auxiliary $SU(2)$ theory.
For reference, we easily compute that:
\begin{eqnarray}
\nn
j_{N_f = 0}(u) &=&
\frac{64 \left(u^2+3\right)^3}{\left(u^2-1\right)^2} \ , \\
\nn
j_{N_f = 1}(u) &=& \frac{16384 \left(3 m-4 u^2\right)^3}{256 m^3-256 m^2 u^2-288 m u+256 u^3+27} \ , \\
\nn &&\\
\nn
j_{N_f = 2}(u) &=& 64 \left(-48 m_1 m_2+64 u^2+3\right){}^3 \big[ -32 m_1^2 \left(m_2^2 \left(3-128
   u^2\right)+2 u \left(64 u^2-9\right)\right)+ \\
\nn
&& \qquad  +16 m_2 m_1 \left(288 m_2^2 u-320
   u^2-3\right)-64 m_2^2 u \left(64 u^2-9\right)+\\
\nn
&& \qquad 
+512 m_1^3 \left(9 m_2 u-8
   m_2^3\right) -432 m_1^4-432 m_2^4+\left(1-64 u^2\right)^2 \big]^{-1} \ , \\
\nn &&
\end{eqnarray}
\begin{eqnarray}\nn
j_{N_f = 3}(u) &=& 8192 \left(\frac{3 m_1^2}{4}+\frac{3 m_2^2}{4}+\frac{3 m_3^2}{4}-12 m_1 m_2 m_3+16
   u^2-u+\frac{1}{256}\right){}^3 \times \\
\nn
&&\quad
\big[ 2 \left(m_1^2-16 m_2 m_3 m_1+m_2^2+m_3^2-2
   u\right){}^3 -
\\ \nn &&\quad
- 864 \left(m_1^2 \left(m_2^2+m_3^2-u\right)+ 
u \left(u-m_3^2\right)+m_2^2
   \left(m_3^2-u\right)\right){}^2 +
\\ \nn &&\quad
+ 9 (64 u+1) \left(-m_1^2+16 m_2 m_3 m_1-m_2^2-m_3^2+2u\right) \times 
\\ \nn &&\quad
\qquad 
\times \left(m_1^2 \left(m_2^2+m_3^2-u\right)+u \left(u-m_3^2\right)+m_2^2
   \left(m_3^2-u\right)\right) -
\\ \nn &&\quad
-32 \left(u+\frac{1}{64}\right)^2 \left(4 (64 u+1)
   \left(m_1^2 \left(m_2^2+m_3^2-u\right)+u \left(u-m_3^2\right)+ 
\right. \right.
\\ \nn &&\quad
\qquad
\left. \left. m_2^2 \left(m_3^2-u\right)\right)
- \left(m_1^2-16 m_2 m_3 m_1+m_2^2+m_3^2-2
   u\right){}^2\right)\big]^{-1} \ .
\\
\end{eqnarray}

Let us now collect the correspondences.
First, the simplicity of the form for $N_f$ does not accommodate any of the groups since there is only one free parameter $u$.
This is not a problem as this case was already matched to the group $\Gamma(2)$ by \cite{malmendier} as mentioned above, which is not an index 24 subgroup.

For $N_f=1$, we find no correspondence for groups Ia and Ib.
However, for the remaining groups, we do find that Seiberg-Witten curves and the elliptic modular surfaces identify so long as we take the following special points in the moduli space of parameters $\{s,u,m_i\}$; for convenience, we shall henceforth take all $m_i = m$ and deal with this equal-mass situation only.
We express the special values as the roots of the following systems of polynomials:
\begin{equation}
\begin{array}{ll}
$IIa$ &: \left\{1088 s^4-16 s^2+1,1024 m^2+64 m+4097,1048576 u^2+8386560 u+16785409\right\}\\
$IIb$ &: \left\{1024 s^4-2048 s^2+1025,1024 m^2+64 m+4097,1048576 u^2+8386560 u+16785409\right\}\\
$IIIa$ &: \left\{4095 s^4-2 s^2-1,32 m-1,1024 u-4097\right\}\\
$IIIb$ &: \left\{4095 s^4-8194 s^2+4095,32 m-1,1024 u-4097\right\}\\
$IIIc$ &: \left\{4096 s^4-1,32 m-1,1024 u-4097\right\} \\
$IIId$ &: \left\{4096 s^4+8192 s^2+4095,32 m-1,1024 u-4097\right\}\\
$IV$ &: \left\{1024 s^6-2048 s^5+1024 s^4+1,\right. \\
& \quad 1048576 m^6+28311552 m^4-393216 m^3+254786560 m^2 -3538880 m+
  764476417, \\
& \quad 1099511627776 u^6+6597069766656 u^5 +16481936998400 u^4+21947284979712 u^3\\
& \qquad \left.
+16445479059456 u^2+6588471429120 u+1105965619201\right\}\\
\end{array}
\end{equation}

For $N_f=2,3$, we can find identification between the modular surfaces and the Seiberg-Witten curves for all our 9 groups, as long as we once again go to special points in the moduli space $\{s,u,m\}$.
The polynomials defining these special points are very involved and we do not present them here.

\begin{figure}[!h!t!b]
\centerline{
(a)
\includegraphics[trim=0mm 0mm 0mm 0mm, clip, width=3.0in]{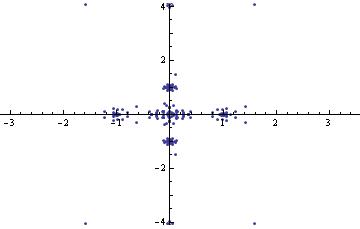}
(b)
\includegraphics[trim=0mm 0mm 0mm 0mm, clip, width=3.0in]{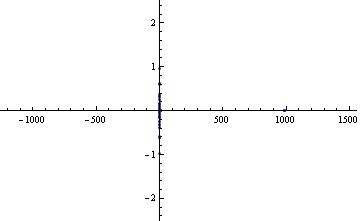}
}
\caption{{\sf {\small
The possible values, on the complex plane, of the $\IP^1_C$ parameter $s$ (drawn in part (a)) and the $u$-plane parameter $u$, which allow for the identification of the modular elliptic surfaces of our 9 index 24 groups and the Seiberg-Witten curves for $SU(2)$ with 0,1,2 or 3 flavours of equal mass and unit cut-off scale.
}}
\label{f:SU2-SW}}
\end{figure}

It is perhaps instructive and certainly amusing to ask which values of the base $\IP^1_C$ coordinate $s$ could we have in order that our modular surfaces can be Seiberg-Witten curves for $SU(2)$ theories.
We find the set of possible solutions for $s$ and plot them en masse on the complex plane, as is drawn in part (a) of \fref{f:SU2-SW}.
For reference, we also find all possible values for $u$, which determines the special values on the so-called $u$-plane of the Seiberg-Witten curves, and include them in part (b) of the figure.
Note that all the allowed values of $u$ are purely imaginary.

\subsection{Enumeration of the Theories}\label{s:count}
Having considered the special index 24 subgroups, in relation to 9 rather special gauge theories, it is natural to wonder how far our prescription in \sref{s:SW-dessin} extends beyond K3 surfaces.
To that effect, it is conducive to our understanding to enumerate all these gauge theories and coset graphs, to which we now take a parting digression.
For our purposes, this is the counting of connected trivalent graphs, which, though constitutes an old and well-developed topic (the reader is referred to some classic works in \cite{cubic}), still is riddled with open conjectures and lack of explicit formulae.

\paragraph{Strictly Cubic Graphs: }
First, let us collect some basic facts.
Since each node is attached to three other nodes (possibly itself), there must be a total of an even number of nodes, say $2n$.
Moreover, let us first consider strictly cubic graphs, namely those without external legs. Thus, we have that the number of edges to be $3n$.
As mentioned throughout the paper, this restriction of not having external legs is the case of all the Schreier-Cayley coset graphs we have considered.
This ``spikeless'' condition is certainly not necessary for the generalized quivers of Gaiotto and we will address them next.

Now, our graphs could be labeled or unlabeled, corresponding to whether an absolute ordering has been assigned or not, respectively, to distinguish the nodes.
It turns out that the unlabeled situation is far more difficult to tackle \cite{wormald}.
Since our nodes are not distinguished in our analyses, we do need to consider this more difficult case\footnote{The labeled case is, of course, also of interest if we keep track of the cosets of the modular group, or of the particular $SU(2)$ gauge group.}.

The majority of work on the subject has been on {\it simple} trivalent graphs, which consists of neither double-edges nor loops (which are arrows going from a node to itself).
In this case, the number of simple cubic graphs, $d_n$, with $2n$ {\it labeled} nodes has explicit recursion relations as given in \cite{wormald}.
The unlabeled case does not enjoy explicit equations, but has been tabulated to quite high values starting at $n=0,1,2,\ldots$,
\begin{equation}
\label{tri-seq}
d_n=
\{
1, 0, 1, 2, 5, 19, 85, 509, 4060, 41301, 510489, 7319447, 117940535, 2094480864,\ldots
\}
\end{equation}
Here, $n=0$ is trivially the null-graph; 
$n=1$ has 2 nodes and clearly has no simple cubic graph:
our sunset graph example in (c) of \fref{f:coset-eg} has multiple edges between the two nodes; $n=2$ has 4 nodes and the single graphs corresponding to $d_2=1$ is isomorphic to the tetrahedron.
Asymptotically, the total number of such simple (not necessarily connected) cubic graphs is \cite{wormald}
\[
d_n \sim \frac{(6n)!}{(3n)!288^n}e^{-2} \ .
\]
This can be considered as the asymptotic number of our gauge theories which have no flavours (i.e., external legs).
Using Stirling's approximation, we see that $d_n \sim \exp((3\log n + \log6 - 3)n-2)$, so that the critical exponent is approximately $(3\log n + \log6 - 3)n$.

Our present concern allows for loops, as can be clearly seen from \fref{f:graph24}, though we do not allow for multiple edges (except for the degenerate case of 2 nodes, corresponding to the ``dumb-bell'').
Now, as mentioned in the discussion surrounding \fref{f:coset-eg}, we have a precise contraction procedure by which one may shrink a triangle to a single node.
Thus, loops can be generated from the simple cubic graphs and the counting is the same as the above, but shifted down in $n$ by 2 since each triangle can be reduced to a loop at a single node.
We can see the graphs explicitly, for example, in the equivalent study of cubic Feynman diagrams \cite{feynman}, particularly in Table IV in cit.~ibid.
The sequence begins with $2,5,19,\ldots$, as in \eqref{tri-seq}, but shifted.
More relevant to our counting is \cite{deMelloKoch:2011uq} since the above trivalent Feynman diagrams distinguish one of the three edges of the vertex.

If one were to ask for the refined counting, distinguishing loops from simple edges,
the best result of which we are aware is the labeled case for a general cubic graph with $s$ simple edges, $d$ double edges and $l$ loops for which explicit recursion formulae are given \cite{chae}.
Since we do not allow multiple edges, $d=0$ except for $n=1$.
Using the notation of \cite{chae}, where $g_1(s,d,l)$ is the number of connected labeled trivalent graphs with $(2n)$ nodes, $s$ simple edges, $d$ double edges and $l$ loops, the recursion of Eqs (14) and (15) give the intertwined relations that
\begin{eqnarray}
\nn
g_1(s,0,l) &=& \frac{2}{l} {2n \choose 2} \left[
(s-2) g_1(s-2,0,l-1) + g_1(s-4,1,l-1) \right] \ ; \\
\nn
g_1(s,1,l) &=&  {2n \choose 2} \left[
2(s-1)g_1(s-1,0,l) + 2g_1(s-3,0,l) + (l+1) g_1(s-2,0,l+1) \right]
 \ , \\
&& g_1(1,0,2)=g_1(6,0,0)=1, \ g_1(3,1,1)=12 , \ g_1(5,2,0)=720 \ ,
\end{eqnarray}
where $l \ge 1$ and, as is clear that the total number of edges must be $3n$, $(s + l) \equiv 0 \mod 3$.

\paragraph{Permitting External Legs: }
If we were to slightly relax our trivalency, and allow also nodes of valency one, some elegant results follow.
This is quite re-assuring, since allowing for valency one is precisely the allowace for external legs and thus will encompass all the relevant gauge theories.
Though, having these ``spikes'' in the coset graphs is more cumbersome to analyze.
Indeed, not permitting external legs is more convenient from the modular group point of view and having them is necessary from the gauge theory side.
Thus, we have included the enumerations for both as a comparison.

It was shown in \cite{vidal} that cubic graphs with external legs and with $n$ nodes (note that $n$ no longer needs to be even), which are unlabeled and connected, proceed with $n$ as
\begin{equation}\label{an-count}
\begin{array}{ll}
a_n = &\{ 1, 1, 2, 2, 1, 8, 6, 7, 14, 27, 26, 80, 133, 170, 348, \\
&765, 1002, 2176, 4682, 6931, 13740, 31085, 48652, 96682, 217152, 
\ldots \}
\end{array}
\end{equation}
One word of caution:
in cit.~ibid.~(cf.~Tab 1-4 on pages 11-13), what Vidal calls ``dotted diagrams'' also include valency two nodes, in addition to the trivalent and univalent external edges.
This is, of course, in line with \fref{f:dessin} where we have inserted a black node on every edge in order to emphasize the ramification at 1728.
One might imagine that this above sequence should be intimately related to the modular group given the trivalent nature of the Cayley graph.
Unsurprisingly, this is indeed so.
The sequence \eqref{an-count} also enumerates the number of conjugacy class of index $n$ subgroups in the modular group.

In fact, Vidal gives the analytic form of the exponential generating function
$D(t) := \sum\limits_{n=0} \frac{a_n}{n!} t^n$ as
\begin{equation}\label{Dt}
D(t) = \sum\limits_{r=1} \frac{\mu(r)}{r} \sum\limits_{k=1} \log
\left(
\sum\limits_{n=0} n! k^n u_{k,n} v_{k,n} t^{rkn} 
\right)
\end{equation}
where $\mu(r)$ is the M\"obius $\mu$-function and 
\begin{equation}
u_{k,n} := \sum\limits_{n_1+2n_2=n} 
        \frac{\chi^{n_1}_{2,k}}{n_1! n_2! k^{n_1+n_2} 2^{n_2}} \ ,
v_{k,n} := \sum\limits_{n_1+3n_2=n} 
        \frac{\chi^{n_1}_{3,k}}{n_1! n_2! k^{n_1+n_2} 3^{n_2}} \ ,
\end{equation}
with the character $\chi_{p,k}$ defined to be $p$ if $k \equiv 0 \mod p$
and 0 otherwise.

Given the form of \eqref{Dt} one can hardly resist finding the plethystics \cite{Benvenuti:2006qr}, which we remind the reader as the following mutual inverse:
\begin{eqnarray}
\nn
&&g(t) = PE[f(t)] = \exp\left(\sum\limits_{k=1}^\infty
  \frac{f(t^k) - f(0)}{k} \right) = {\prod\limits_{n=1}^\infty
    (1-t^n)^{-b_n}} \\
&\Leftrightarrow&
f(t) = PE^{-1}(g(t)) = \sum_{r=1}^\infty  \frac{\mu(r)}{r} \log
  (g(t^r)) \ ,
\end{eqnarray}
where $f(t) = \sum\limits_{n=0}^\infty b_n t^n$ is the Taylor series for $f(t)$.A theme in \cite{Benvenuti:2006qr} was to see whether $f(t)$ becomes a rational function and can serve as the Hilbert series of the vacuum moduli space of a supersymmetric gauge theory.
Here, therefore, we have that $g(t)$, the generating function for the possibly disconnected diagrams of our concern, being
$g(t) = \prod\limits_{k=1}^\infty \sum\limits_{n=1}^\infty n! k^n u_{k,n} v_{k,n} t^{kn}$.

\newpage

\section*{Acknowledgements}
We gratefully acknowledge many helpful comments by
Alessio Corti, Amihay Hanany, Sanjaye Ramgoolam,
Diego Rodrigez-Gomez and Ed Segal.
YHH would like to thank the Science and
Technology Facilities Council, UK, for an Advanced Fellowship and for STFC 
grant ST/J00037X/1, the Chinese Ministry of Education, for a Chang-Jiang
Chair Professorship at NanKai University, the US NSF for grant CCF-1048082, 
as well as City University, London and Merton College, Oxford,
for their enduring support.
JM is grateful to the NSERC of Canada.



\begin{thebibliography}{99}
\bibitem{classSebbar}
A.~Sebbar, 
``Classification of torsion-free genus zero congruence groups'', 
Proc. Amer. Math. Soc. 129 (2001), 2517--2527.


\bibitem{mckaysebbar}
J.~McKay and Abdellah Sebbar, ``Arithmetic Semistable Elliptic Surfaces'', 
Proceedings on Moonshine and related topics (Montr\'eal, QC, 1999), 119--130,
CRM Proc. Lecture Notes, 30, Amer. Math. Soc., Providence, RI, 2001.

\bibitem{sebbar}
Abdellah Sebbar, 
``Modular subgroups, forms, curves and surfaces'', 
Canad. Math. Bull. 45 (2002), no. 2, 294--308.

\bibitem{Gaiotto:2009we}
  D.~Gaiotto,
  ``N=2 dualities,''
  [arXiv:0904.2715 [hep-th]].

\bibitem{Hanany:2010qu}
  A.~Hanany, N.~Mekareeya,
  ``Tri-vertices and SU(2)'s,''
  JHEP {\bf 1102}, 069 (2011).
  [arXiv:1012.2119 [hep-th]].

\bibitem{Feng:2005gw}
  B.~Feng, Y.~-H.~He, K.~D.~Kennaway, C.~Vafa,
  ``Dimer models from mirror symmetry and quivering amoebae,''
  Adv.\ Theor.\ Math.\ Phys.\  {\bf 12}, 3 (2008).
  [hep-th/0511287].

\bibitem{CayleyGamma}
Javier Aramayona, ``Aspects of the geometry of mapping class groups'', in
Disertaciones del Seminario de Matem\'aticas Fundamentalies, 37,
(2006) 
\verb|http://dmle.cindoc.csic.es/pdf/DISERTACIONESMATEMATICAS_37.pdf|

\bibitem{serre}
J-P~Serre, ``Cours d'arithm\'etique,'' 
Presses Universitaires de France; Paris, 1970.

\bibitem{Milne}
J.~S.~Milne, ``Modular Functions and Modular Forms,''
2009, Available at \verb|www.jmilne.org/math/|

\bibitem{cox}
D.~A.~Cox, J.~McKay, and P.~Stevenhagen, ``Principal Moduli and Class Fields''. Bull. Lond. Math. Soc. 2004, 36 (1): 3-12

\bibitem{TopYui}
Jaap Top and Noriko Yui, ``Explicit Equations of Some Elliptic Modular 
Surfaces'', Rocky Mountain J. Math. Volume 37, Number 2 (2007), 663-687.

\bibitem{ADE}
John McKay, ``Graphs, singularities and finite groups'',
Proc. Symp. Pure Math. (Amer. Math. Soc.) (1980) 37: 183 and 265 


\bibitem{shioda}
T.~Shioda, ``On elliptic modular surfaces'', J.~Math.~Soc.~Japan 24 (1972),
20-59.

\bibitem{shioda-rev}
Matthias Schuett and Tetsuji Shioda, ``Elliptic Surfaces'',
arXiv:0907.0298v3.


\bibitem{LY}
Ron Livn\'e and Noriko Yui,
``The modularity of certain non-rigid Calabi-Yau threefolds'',
arXiv:math/0304497.


\bibitem{MP}
R.~Miranda, U.~Persson, ``Configurations of In Fibers on Elliptic K3 surfaces'',Math. Z. 201 (1989), 339361.


\bibitem{hep-th/0611082} 
  S.~K.~Ashok, F.~Cachazo and E.~Dell'Aquila,
  ``Children's drawings from Seiberg-Witten curves,''
  hep-th/0611082.




\bibitem{Jejjala:2010vb}
  V.~Jejjala, S.~Ramgoolam, D.~Rodriguez-Gomez,
  ``Toric CFTs, Permutation Triples and Belyi Pairs,''
  JHEP {\bf 1103}, 065 (2011).
  [arXiv:1012.2351 [hep-th]].

\bibitem{Hanany:2011ra}
  A.~Hanany, Y.~-H.~He, V.~Jejjala, J.~Pasukonis, S.~Ramgoolam, D.~Rodriguez-Gomez,
  ``The Beta Ansatz: A Tale of Two Complex Structures,''
  JHEP {\bf 1106}, 056 (2011).
  [arXiv:1104.5490 [hep-th]].


\bibitem{Hanany:2005ve}
  A.~Hanany, K.~D.~Kennaway,
  ``Dimer models and toric diagrams,''
  [hep-th/0503149].

\bibitem{BM}
F.~Beukers and H.~Montanus, ``Explicit calculation of elliptic fibrations of K3-surfaces and their Belyi-maps'', Number theory and polynomials, 3351, London Math. Soc. Lect. Note Ser. 352, Cambridge Univ. Press, Cambridge, 2008.
\verb|http://www.staff.science.uu.nl/~beuke106/mirandapersson/Dessins.html|

\bibitem{He:2010mh}
  Y.~-H.~He,
  ``On Fields over Fields,''
  [arXiv:1003.2986 [hep-th]].



\bibitem{Meyer}
Christian Meyer, ``Modular Calabi-Yau Threefolds'',
Fields Institute Monographs, vol. 22, ISBN-13: 978-0-8218-3908-9.

\bibitem{Livne}
R.~Livne, ``Motivic Orthogonal Two-dimensional Representations of Gal(Q=Q)'', 
Israel J. of Math. 92 (1995), 149-156.

\bibitem{Schuett}
Matthias Schuett , ``Arithmetic of K3 surfaces'', arXiv:0809.0830.\\
--, PhD dessertation, available at\\
\verb|http://www.iag.uni-hannover.de/~schuett/Dissertation.pdf|


\bibitem{shimada}
I.~Shimada, D.-Q.~Zhang, 
``Classification of extremal elliptic K3 surfaces and fundamental groups of open K3 surfaces'', arXiv:math/0007171.

\bibitem{Hanany:1998sd}
  A.~Hanany, Y.~-H.~He,
  ``NonAbelian finite gauge theories,''
  JHEP {\bf 9902}, 013 (1999).
  [hep-th/9811183].

\bibitem{Seiberg:1994rs}
  N.~Seiberg, E.~Witten,
  ``Electric - magnetic duality, monopole condensation, and confinement in N=2 supersymmetric Yang-Mills theory,''
  Nucl.\ Phys.\  {\bf B426}, 19-52 (1994).
  [hep-th/9407087].

\bibitem{Witten:1997sc}
  E.~Witten,
  ``Solutions of four-dimensional field theories via M theory,''
  Nucl.\ Phys.\  {\bf B500}, 3-42 (1997).
  [hep-th/9703166].


\bibitem{Tai:2010im}
  T.~S.~Tai,
  ``Triality in SU(2) Seiberg-Witten theory and Gauss hypergeometric function,''
  Phys.\ Rev.\  D {\bf 82}, 105007 (2010)
  [arXiv:1006.0471 [hep-th]].

\bibitem{Alday:2009aq}
  L.~F.~Alday, D.~Gaiotto and Y.~Tachikawa,
  ``Liouville Correlation Functions from Four-dimensional Gauge Theories,''
  Lett.\ Math.\ Phys.\  {\bf 91}, 167 (2010)
  [arXiv:0906.3219 [hep-th]].

\bibitem{Eguchi:1997pu}
  T.~Eguchi,
  ``Seiberg-Witten theory and S-duality,''
{\it NATO Advanced Study Institute on Strings, Branes and Dualities, Cargese, France, 26 May - 14 Jun 1997}


\bibitem{patterson}
D. B. Patterson. The fundamental group of the modulus space. Michigan
Math. J., 26(2):213223, 1979.
\comment{
G.~Gonz\`alez Diez and P.~Lochark, ``On the fundamental groups at infinity of the moduli spaces of compact Riemann surfaces'',
Michigan Math. J. 49 (2001), 493-500.
}


\bibitem{malmendier}
Andreas Malmendier, ``Kummer surfaces associated with Seiberg-Witten curves'',
J. Geom. Phys. 62 (2011), no. 1, 107-123, ArXiv:0912.4774.


\bibitem{cubic}
R.~C.~Read, ``Some enumeration problems in graph theory ``. 
Doctoral thesis, University of London,1958.\\
R.~W.~Robinson, ``Counting cubic graphs'', J.~Graph Theory, 1 (1977), 285-286.\\R.~W.~Robinson and N.~C.~Wormald, ``Numbers of Cubic Graphs'', Journal of Graph Theory, Vol. 7 (1983) 463-467.

\bibitem{wormald}
N.~C.~Wormald, ``Enumeration of labelled graphs II: Cubic graphs with a given connectivity'', J.~London Math.~Soc.(2), 20 (1979).

\bibitem{feynman}
Hagen Kleinert, Axel Pelster, Boris Kastening, and Michael Bachmann,
``Recursive graphical construction of Feynman diagrams and their multiplicities
in $\phi^4$ and $\phi^2A$ theory'', Phys.~Rev.~E, Vol.~62, Num.~2, 2000.

\bibitem{deMelloKoch:2011uq} 
  R.~de Mello Koch and S.~Ramgoolam,
  ``Strings from Feynman Graph counting : without large N,''
  Phys.\ Rev.\ D {\bf 85}, 026007 (2012)
  [arXiv:1110.4858 [hep-th]].


\bibitem{chae}
Gab-Byung Chae, Edgar M.~Palmer, Robert W.~Robinson, ``Counting labeled general cubic graphs'', Discrete Mathematics 307 (2007).

\bibitem{vidal}
S.~A.~Vidal, ``Sur la Classification et le Denombrement des Sous-groupes du Groupe Modulaire et de leurs Classes de Conjugaison,'' ArXiv:math/0702223.


\bibitem{Benvenuti:2006qr} 
  S.~Benvenuti, B.~Feng, A.~Hanany and Y.~-H.~He,
  ``Counting BPS Operators in Gauge Theories: Quivers, Syzygies and Plethystics,''
  JHEP {\bf 0711}, 050 (2007)
  [hep-th/0608050].
\end{thebibliography}
\end{document}